\documentclass[%reprint,
amsmath,
twocolumn,
]{revtex4-2}

\usepackage{graphicx} % Include figure files
\usepackage{hyperref} % add hypertext capabilities
\usepackage{multirow}

\usepackage[usenames]{xcolor}

\newcommand \figfile[1]{#1}

\begin{document}

%\preprint{Cockcroft-2011-07}

\title{Transverse phase space tomography in the CLARA accelerator test facility using image compression and machine learning}% Force line breaks with \\
%\thanks{Work supported by the Science and Technology Facilities Council, UK;}%

\author{A.\,Wolski}
\email{a.wolski@liverpool.ac.uk}
\affiliation{University of Liverpool, Liverpool, UK, and the Cockcroft Institute, Daresbury, UK.}

\author{M.A.\,Johnson}
\affiliation{STFC/ASTeC, Daresbury Laboratory, Daresbury, UK.}

\author{M.\,King}
\affiliation{STFC/ASTeC, Daresbury Laboratory, Daresbury, UK.}

\author{B.L.\,Militsyn}
\affiliation{STFC/ASTeC, Daresbury Laboratory, Daresbury, UK.}

\author{P.H.\,Williams}
\affiliation{STFC/ASTeC, Daresbury Laboratory, Daresbury, UK.}

\date{\today}% It is always \today, today,

\begin{abstract}
We describe a novel technique, based on image compression and machine learning, for transverse phase
space tomography in two degrees of freedom in an accelerator beamline.  The technique has been
used in the CLARA accelerator test facility at Daresbury Laboratory: results from the machine learning
method are compared with those from a conventional tomography algorithm (algebraic reconstruction),
applied to the same data.  The use of machine learning allows reconstruction of the 4D phase space
distribution of the beam to be carried out much more rapidly than using conventional tomography
algorithms, and also enables the use of image compression to reduce significantly the size of the data
sets involved in the analysis.  Results from the machine learning technique are at least as good as those
from the algebraic reconstruction tomography in characterising the beam behaviour, in terms of the variation
of the beam size in response to variation of the quadrupole strengths.
\end{abstract}

%\pacs{29.27.Bd,29.27.Fh,41.75.Ht,41.85.Ew}
\maketitle

%\tableofcontents

% ------------------------------------------------------------------------------

\section{Introduction\label{sec:intro}}

% ------------------------------------------------------------------------------

Phase space tomography provides a valuable technique for understanding the
properties of a beam in a particle accelerator, and has been applied in a range of
different machines, for example \cite{mckee1995, yakimenko2003,stratakis2003,
stratakis2007, xiang2009,rohrs2009,xing2018,ji2019}. However, conventional
tomography techniques present some challenges, including the presence of artefacts
in the reconstruction (which can be especially prominent when the number of projections
is limited), and the computational time and resources required to construct the phase
space distribution with good resolution.  Tomography in two transverse degrees of
freedom allows characterisation of betatron coupling, but the sizes of the data structures
required for the analysis increase rapidly with the dimensionality of the system. Storage
of a 4D phase-space distribution in an array with dimension $N$ along each axis requires
a data structure of $N^4$ numerical values, and the memory resources needed while
processing the input data to construct the phase space can be much larger.  The
demands on computing power increase rapidly with increasing dimensionality of the phase
space, and this may limit the use of high-dimensional phase space tomography (with good
resolution) in applications where it could make a valuable contribution to machine
operation, for example in short-pulse, short-wavelength free electron lasers
\cite{alesini2006} or injectors for machines using novel acceleration technologies such
as plasma cells or dielectric wakefield structures \cite{marx2019, jastermerz2022}.

Recent work \cite{jastermerz2022}
has shown (in simulation) how phase space tomography can be performed in
$2\frac{1}{2}$ degrees of freedom, to provide transverse phase space properties
as a function of longitudinal position along a bunch. Steps have been taken towards
full 6D phase space tomography, but the methods that have been employed (which
include the use of machine learning) have not so far allowed the full reconstruction
of the 6D phase space \cite{scheinker2022}.  Where betatron coupling or
synchro-betatron coupling are present, tracking a beam from a given point in the
accelerator to determine its properties as function of position in the beamline requires
the full phase space in the coupled degrees of freedom to be described, and in complex
machines where multiple correlations can be present, full 6D phase space reconstruction
would provide all the necessary information. Techniques allowing reduction
of the processing time and data storage requirements for high-dimensional phase
space tomography offer the prospect of enabling routine complete and detailed
characterization of the charge distribution within bunches in an accelerator, including
all cross-plane correlations, with significant benefits for advanced accelerator facilities.

In principle, image compression techniques can be used to reduce the size of the
data structures needed to store and process tomography data while maintaining the
potential for reconstructing the phase space with a given resolution.  Reduction in the
size of the data sets can also be accompanied by reduction in the time taken to process
those data sets.  However, it is not clear how existing tomography algorithms can be
adapted so that they can be applied directly to compressed data.  Machine learning
techniques offer an alternative to conventional tomography methods, and have the
potential to allow direct tomographic analysis of data in a compressed form.  Machine
learning is already extensively used for image analysis and tomography, particularly in
medical contexts \cite{wang2020}. There is also increasing interest in the use of machine
learning for a range of applications in accelerator design and operation, including design
optimization \cite{licheng2018, wan2019, edelen2020}, modelling \cite{emma2018},
collection and analysis of diagnostic data \cite{azzopardi2019, xu2020, tennant2020,
omarov2022}, and operational optimization \cite{emery2021, arpaia2021}.  Recent work
\cite{scheinker2022} has shown (in simulation) the use of a neural network for constructing
two-dimensional projections of a six-dimensional phase space.

In the current paper, we report results of
experimental studies aimed at demonstrating the use of machine learning for
phase space tomography, working with beam images and phase-space distributions
stored in compressed form.  We describe the principles of the technique, compare
the results with those using a conventional tomography algorithm on the same data
sets, and discuss the potential advantages of the use of machine learning for this
application.

The experimental work that we present has been carried out on CLARA, the
Compact Linear Accelerator for Research and Applications at Daresbury Laboratory
\cite{claracdr,clara1,angalkalinin2020}.  Relevant features of the facility are outlined
in Section~\ref{sec:tomography2dof}, in which we also describe the experimental
technique (Section~\ref{subsec:experimentaltechnique}), and present the results
of an analysis of the experimental data using a conventional tomography algorithm,
algebraic reconstruction
(Section~\ref{subsection:artanalysis}).  In Section~\ref{sec:ml} we describe and present
results from the tomography analysis based on machine learning.  Some conclusions
from the work are discussed in Section~\ref{sec:conclusions}.

% ------------------------------------------------------------------------------

\section{Characterization of transverse phase space in CLARA using a conventional tomography technique\label{sec:tomography2dof}}

% ------------------------------------------------------------------------------

Previous studies of phase-space tomography in two transverse degrees of freedom using
CLARA were reported in \cite{wolski2020}.
At the time of the previous tomography studies, carried out in 2019, the facility (CLARA Front
End), included an electron source and short linac designed to provide bunches at a repetition
rate of 10\,Hz with charge up to 250\,pC, momentum up to 50\,MeV/c, and transverse emittance
below 1\,{\textmu}m.  Because of technical limitations during the tomography data collection,
measurements in 2019 were made with beam momentum 30\,MeV/c, and bunch charge up to
50\,pC.  Since then, CLARA has undergone further development, with a number of changes to
components and layout; however, the recent measurements reported here were made with
parameters comparable to those used in the original study, specifically with beam momentum
35\,MeV/c, and bunch charge up to 100\,pC.  Further development of CLARA is planned, both to
extend the energy reach, and to test new RF gun technology, in particular a low-emittance high
repetition-rate source (HRRG).  Detailed characterisation of the HRRG performance will include
studies of the transverse phase-space.  Work to develop novel phase-space
tomography techniques, in particular making use of image compression and machine learning, has
been motivated by the need to facilitate beam characterisation in CLARA generally, and HRRG
performance in particular.  The results reported here are from recent measurements on CLARA
in its current form, with the existing 10\,Hz RF electron gun.

\subsection{Experimental technique: design parameters and calibrated model\label{subsec:experimentaltechnique}}
The tomography technique described in \cite{wolski2020} was applied to CLARA, following upgrade
work performed since the previous tomography studies. Some changes were made to details of the experimental
procedure to take account of changes in the beam optics and machine layout; however, the overall
procedure remained the same in its essential points.  A beam momentum of 35\,MeV/c was used.
Measurements were made using a section
of beamline between the exit of the linac (the ``reconstruction point'') and a fluorescent screen
on which the transverse beam profile could be observed (the ``observation point'').  The beamline
between the reconstruction point and the observation point contains five quadrupoles.

 \begin{figure*}
 \includegraphics[trim = 60pt 90pt 250pt 30pt, clip, width=2\columnwidth,]{\figfile{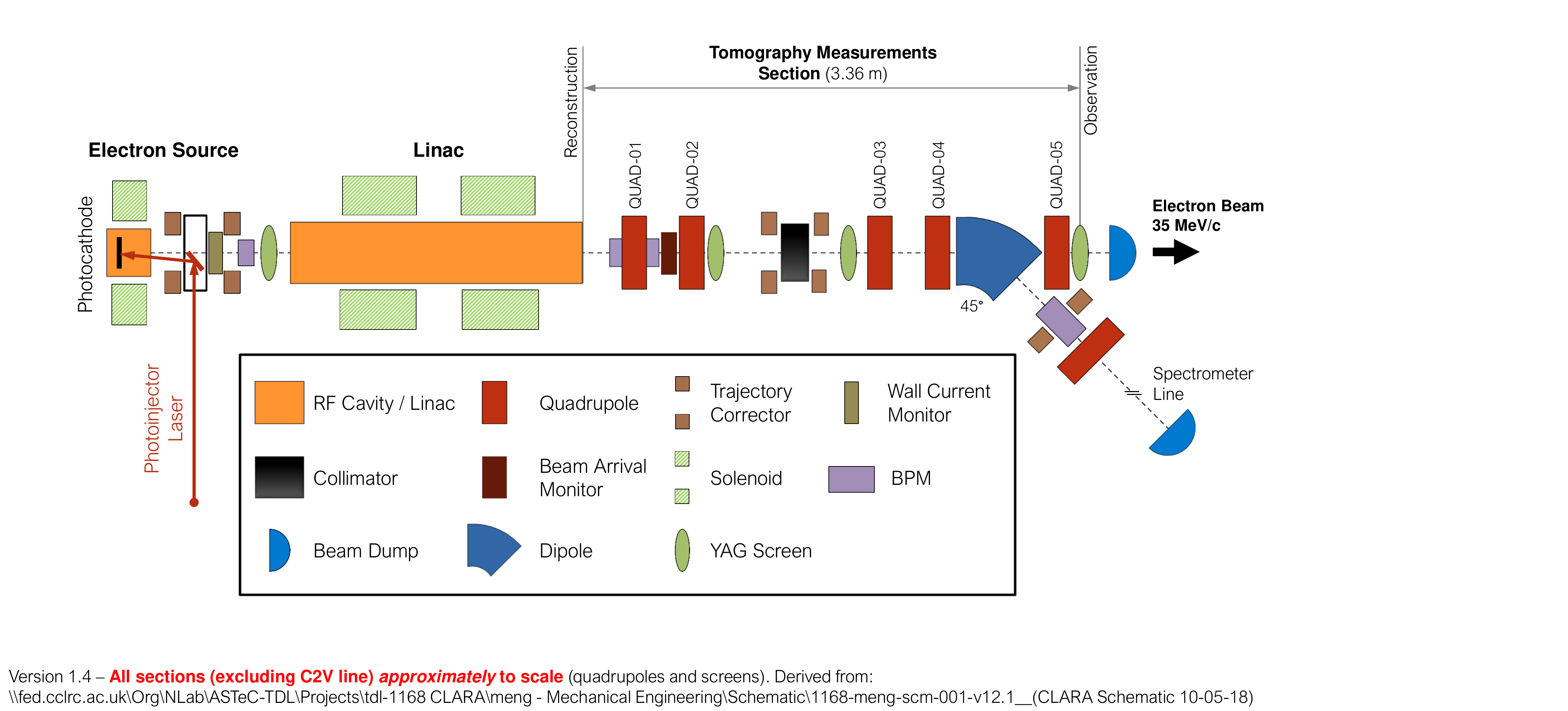}}
 \caption{Layout of CLARA, showing the electron source, magnetic elements, linac and diagnostics.  Distances
between elements are shown approximately to scale. The optical functions and phase space distribution at the
exit of the linac (the reconstruction point) were calculated from the measured beam profiles on the third YAG
screen after the linac (the observation point) for different strengths of the five quadrupoles, QUAD-01 to QUAD-05.
 \label{claralayoutschematic}}
 \end{figure*}

To prepare for the measurements, a machine model \cite{scott2019} was used to determine gradients for the
five quadrupoles in the measurement section that would allow control of the betatron phase advances
between the reconstruction and observation points, while keeping approximately constant the beta
functions at the observation point (see the schematic layout of CLARA in Fig.~\ref{claralayoutschematic}).
A sequence of 32 sets of quadrupole gradients was determined, providing a good range of variation in
horizontal and vertical betatron phase advance over the sequence.  Maintaining constant, and approximately
equal beta functions at the observation point helps to provide good conditions for beam profile measurements:
if the beam image has a large aspect ratio, or gets too large or too small, it can be difficult to determine
accurately the beam sizes.  Data collection consisted of recording the beam profile for each of the 32 steps
in the sequence.  The order of steps in the sequence was chosen to minimise the changes in strength of
the magnets from each step to the next, and in particular to avoid as far as possible changes in polarity:
this helps to reduce the frequency with which the magnets need to be degaussed (the quadrupoles were
degaussed at the start of each scan, and midway through the scan).  At each step, ten screen images
were recorded, plus an extra image with the photoinjector laser turned off to allow for  subtraction of
background resulting from dark current.  A complete quadrupole scan was carried out first with bunch charge
10\,pC, and then with bunch charge 100\,pC.  Although space-charge effects in the injector are significant
at 100\,pC, in the measurements section at beam momentum 35\,MeV/c space-charge has little impact.

The analysis presented here is carried out in normalised phase space: this helps to improve the
accuracy of the phase space reconstruction \cite{hock2011}. Since the section of beamline in CLARA
where the measurements were performed consists only of drift spaces and normal quadrupoles,
coupling can be neglected in constructing the normalising transformation; however, it should be
emphasised that the data analysis nevertheless still allows for full characterisation of any coupling in 
the beam.  Normalised horizontal phase space co-ordinates $(x_N, p_{xN})$ at a
particular location along the beamline are related to the physical co-ordinates $(x,p_x)$ by:
\begin{equation}
\left( \begin{array}{c}
x_N \\ p_{xN}
\end{array} \right) = 
\left( \begin{array}{cc}
\frac{1}{\sqrt{\beta_x}} & 0 \\
\frac{\alpha_x}{\sqrt{\beta_x}} & \sqrt{\beta_x}
\end{array} \right)
\left( \begin{array}{c}
x \\ p_{x}
\end{array} \right),
\label{normalisingtransformation}
\end{equation}
where $\alpha_x$, $\beta_x$ are the usual Courant--Snyder optics functions at the specified
beamline location.  If the phase space distribution is matched to the optics functions, then the
distribution in normalised co-ordinates $\rho_N(x_N, p_{xN})$ will be invariant
under rotations in phase space.  Furthermore, the transport matrices in normalised phase space
are simply rotation matrices (through angles corresponding to the phase advance), so a matched
phase space distribution will be invariant under linear transport along the beamline.

In practice, the phase space distribution is not known in advance: the goal of the measurement
is to determine the distribution.  Phase space normalisation cannot, therefore, be carried out using
optics functions known to be exactly matched to the phase space distribution.
Instead, a machine model is used to generate an expected distribution, and the optics functions
describing this distribution are used to normalise the phase space.  If the real beam distribution
is reasonably close to that expected from the machine model, then in the normalised phase space
the real beam distribution will have at least approximate rotational symmetry.  Phase space
tomography (in normalised phase space) can be used to determine the actual distribution, which
can be transformed back to the physical co-ordinates using the inverse of the normalising
transformation given in Eq.~(\ref{normalisingtransformation}).

For the measurements in CLARA, a design model of the machine was used to predict the phase
space beam distribution at the reconstruction point (the exit of the linac: see
Fig.~\ref{claralayoutschematic}).  The values of the optics functions are shown in
Table \ref{machineparameterstable}.
Preliminary analysis of the experimental data was carried out using the parameter values
from the design model.  The results indicated substantial differences between the design values
and the real values, largely arising from differences between the operational settings actually used
for the injector and linac, and the settings assumed in the machine model when preparing for the
experiments.  Furthermore, closer investigations found that the magnetic lengths of the
quadrupoles in the beamline following the linac (the section used for the tomography studies)
were somewhat larger than had been thought, resulting in changes in the transfer matrices between the
reconstruction point and the observation point for the quadrupole gradients (calculated using the
design model) used during the quadrupole scan.

\begin{table}
\caption{Parameter values in the design and calibrated CLARA machine models.  Optics functions
are given at the tomography reconstruction point (the exit of the linac).\label{machineparameterstable}}
\begin{tabular}{ccc}
\hline
Parameter & Design model & Calibrated model \\
\hline
$\beta_x$ & 15.0\,m & 5.5\,m \\
$\alpha_x$ & -3.4 & 0.0 \\
% horizontal emittance $\gamma\varepsilon_x$ & 0.52\,$\mu$m & - \\
% vertical emittance $\gamma\varepsilon_y$ & 0.52\,$\mu$m & - \\
$\beta_y$ & 15.0\,m & 5.5\,m \\
$\alpha_y$ & -3.4 & 1.5 \\
quadrupole mag. length & 100.7\,mm & 127.0\,mm \\
\hline
\end{tabular}
\end{table}

\begin{figure}
 \includegraphics[trim = 50pt 10pt 100pt 70pt, clip, width=\columnwidth,]{\figfile{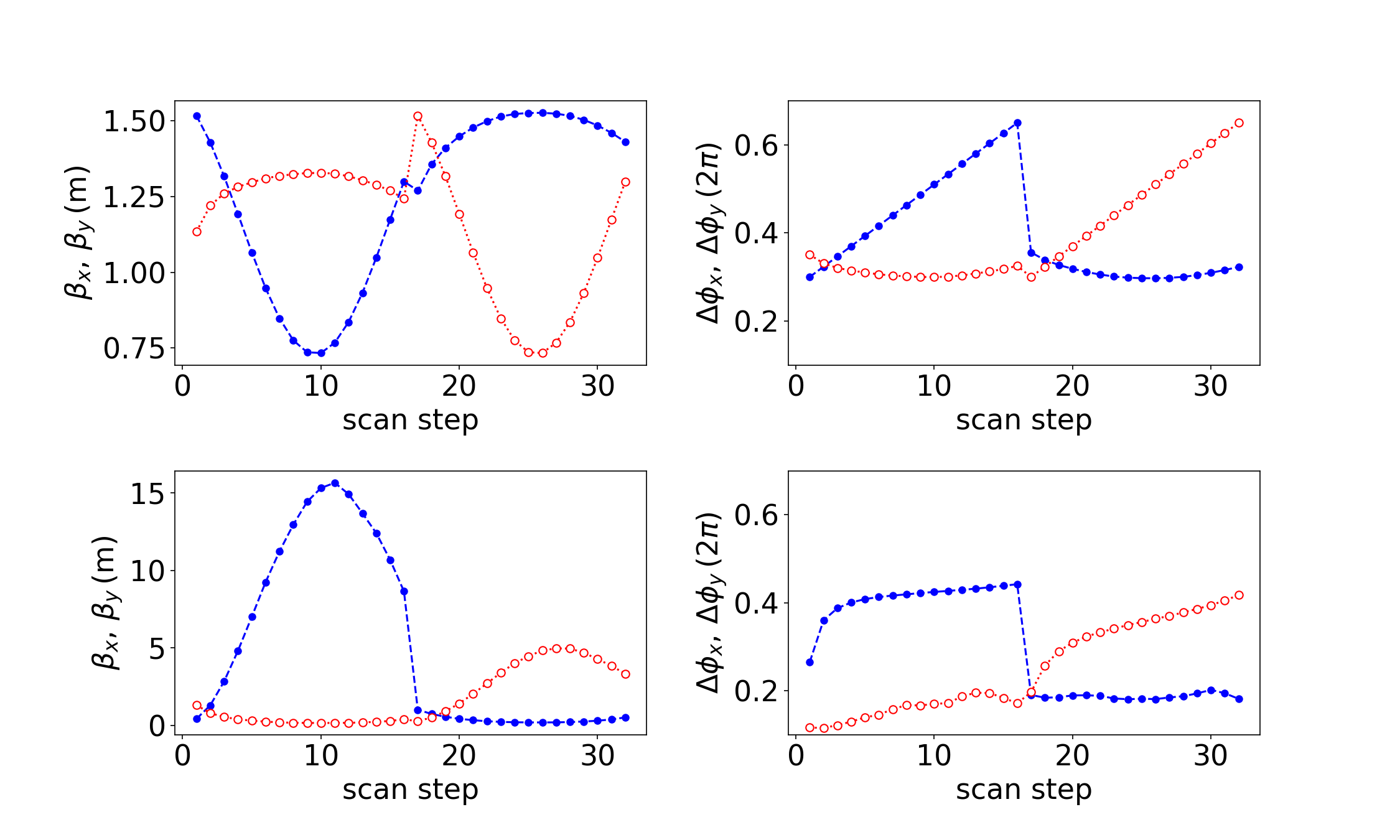}}
 \caption{Optics functions in the CLARA tomography measurements, in the design model (top) and in the
 calibrated model (bottom).  Left-hand plots show the beta functions at the observation point at each
 step in the quadrupole scan; right-hand plots show the phase advances from the reconstruction point to
 the observation point at each step in the quadrupole scan.  Blue solid points (with dashed lines) show the
 values in the horizontal plane; red open points (with dotted lines) show the values in the vertical plane.
 \label{opticsfunctionsdesignandcalibrated}}
 \end{figure}

Differences between the design parameters and the calibrated model are evident in
Fig.~\ref{opticsfunctionsdesignandcalibrated}, which shows the beta functions at the observation
point and the phase advances from reconstruction to observation point, at each step in the quadrupole
scan using the design quadrupole gradients.  Note that the steps were not followed in the order
in Fig.~\ref{opticsfunctionsdesignandcalibrated}, which shows the steps in order of increasing
horizontal phase advance, followed by increasing vertical phase advance: as mentioned above, the
actual order of the steps during the measurements was designed to minimise the changes in quadrupole
strengths between successive steps, to reduce the need for degaussing. The quadrupole gradients used
in the scan were determined using the
design model (top plots in Fig.~\ref{opticsfunctionsdesignandcalibrated}); the same gradients, when
used in the calibrated model with the revised quadrupole lengths and optics functions, lead to the
observation point beta functions and phase advances shown in the bottom plots in
Fig.~\ref{opticsfunctionsdesignandcalibrated}.  Following the initial analysis of the quadrupole scan
data using the design parameters, the analysis was repeated using the parameters for the calibrated
model (and the transfer matrices calculated using the design quadrupole gradients).  The optics for
the design model are shown in Fig.~\ref{opticsfunctionsdesignandcalibrated} only to illustrate the
intended conditions for the tomography data collection, and for comparison with those for the
calibrated model.  In the remainder of this work, we refer only to the calibrated model.

\subsection{Quadrupole scan analysis using the algebraic reconstruction tomography technique\label{subsection:artanalysis}}
Screen images collected during the quadrupole scans were used in
an algebraic reconstruction tomography (ART) code, to determine the 4D transverse phase space
charge distribution.  The same tomography code was used for the recent data as was used in the
studies on CLARA FE: the earlier work included validation of the code, using simulated data \cite{wolski2020}.
In principle, since the only changes in machine settings made during the course of a quadrupole scan
are to the quadrupole gradients, the phase space distribution at the reconstruction point (in the current
studies, at the exit of the linac, upstream of the quadrupoles) should vary little during a
scan.

Beam images collected during a quadrupole scan are prepared for the tomography analysis by first
subtracting a background image (to remove any artefacts from dark current), and then cropping
and scaling the images.  To crop the images, we remove the area outside a certain range of pixels from
the point of peak intensity in the image. The same cropping range is used on each step in the quadrupole
scan, so that the cropped images all have the same dimensions in pixels.  The
crop limits are chosen so that the beam occupies as much of the cropped images as possible, without
clipping the beam in any of the images.  To scale the images, we demagnify each image along
each axis by the square root of the beta function corresponding to that axis (while maintaining the same
number of pixels in each image).  In effect, scaling means that given an initial calibration factor in mm/pixel,
the calibration factor after scaling will be in mm/$\sqrt{\mathrm{m}}$/pixel.  The beta functions used
for scaling are found from the optics in the calibrated model (propagating the values from the reconstruction
point to the observation point, using the transfer matrix calculated from the corresponding quadrupole
strengths).  Scaling essentially transforms the images to normalised phase space: this means that
if the phase space distribution at the reconstruction point was correctly matched to the optics in the
calibrated model, then the scaled beam size (in pixels) would remain constant over the course of the
quadrupole scan.  Finally, the resolution of the normalised images is reduced (or increased, if
necessary) to 39$\times$39 pixels.

For the tomography analysis (using ART), we reconstruct the 4D phase space with a resolution equal
(in pixels) to the image resolution, i.e.~39 pixels on each axis.  The phase space resolution is not in principle
constrained by the technique, but is a practical choice, decided by a balance between the desired level of detail
in the reconstructed phase space distribution, and the computation time and resources needed for the
analysis (which can increase rapidly with increasing phase space resolution).
The results of the tomography can be validated by transporting, for each step in the
quadrupole scan, the 4D phase space distribution from the reconstruction point to the observation point
using the transfer matrix calculated from the known quadrupole strengths and drift lengths; and then
comparing the projection onto co-ordinate space with the corresponding observed beam image.
 
\begin{figure*}
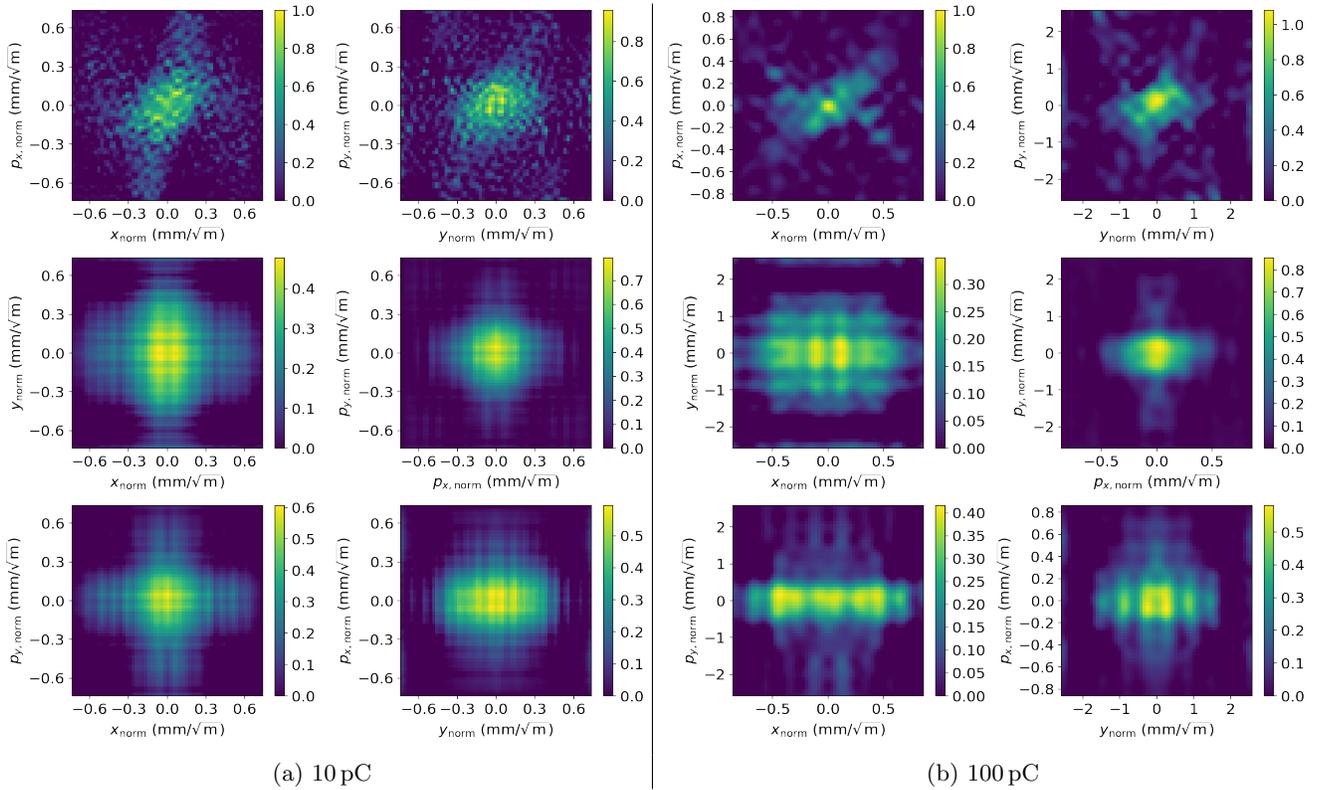

 \begin{tabular}{c|c}
 \includegraphics[trim = 0pt 35pt 65pt 80pt, clip, width=\columnwidth,]{\figfile{PS4DPhaseSpaceProjections-ART-10pC}} &
 \includegraphics[trim = 0pt 35pt 65pt 80pt, clip, width=\columnwidth,]{\figfile{PS4DPhaseSpaceProjections-ART-100pC}} \\
  (a) 10\,pC &
  (b) 100\,pC
 \end{tabular}
 \caption{Projections of the 4D phase space distribution of the beam in CLARA at the  exit of the linac, for
 (a) 10\,pC bunch charge and (b) 100\,pC bunch charge, found from algebraic reconstruction tomography.
 \label{4dphasespaceart}}
 \end{figure*}

\begin{figure*}
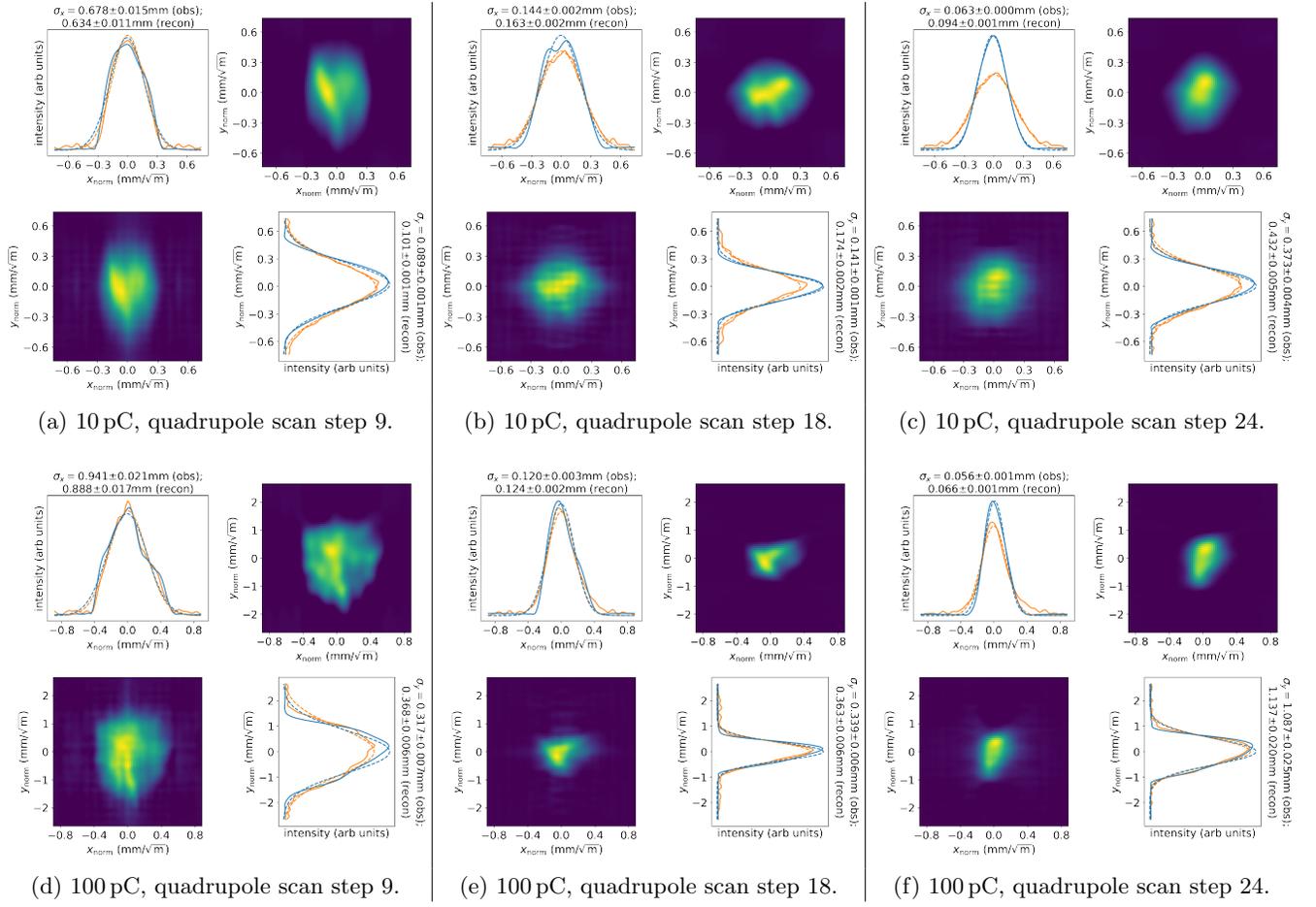

 \begin{tabular}{c|c|c}
 \includegraphics[trim = 25pt 35pt 60pt 60pt, clip, width=0.67\columnwidth,]{\figfile{PS4DTomographySinograms-ART-Resn39-09-10pC}} &
 \includegraphics[trim = 25pt 35pt 60pt 60pt, clip, width=0.67\columnwidth,]{\figfile{PS4DTomographySinograms-ART-Resn39-18-10pC}} &
 \includegraphics[trim = 25pt 35pt 60pt 60pt, clip, width=0.67\columnwidth,]{\figfile{PS4DTomographySinograms-ART-Resn39-24-10pC}} \\
 (a) 10\,pC, quadrupole scan step 9. &
 (b) 10\,pC, quadrupole scan step 18. &
 (c) 10\,pC, quadrupole scan step 24. \\[12pt]
 \includegraphics[trim = 25pt 35pt 60pt 60pt, clip, width=0.67\columnwidth,]{\figfile{PS4DTomographySinograms-ART-Resn39-09-100pC}} &
 \includegraphics[trim = 25pt 35pt 60pt 60pt, clip, width=0.67\columnwidth,]{\figfile{PS4DTomographySinograms-ART-Resn39-18-100pC}} &
 \includegraphics[trim = 25pt 35pt 60pt 60pt, clip, width=0.67\columnwidth,]{\figfile{PS4DTomographySinograms-ART-Resn39-24-100pC}} \\
 (d) 100\,pC, quadrupole scan step 9. &
 (e) 100\,pC, quadrupole scan step 18. &
 (f) 100\,pC, quadrupole scan step 24.
 \end{tabular}
 \caption{Validation images from algebraic reconstruction tomography, from three steps in the quadrupole scan
 at two different bunch charges: 10\,pC in cases (a), (b), (c); and 100\,pC in cases (d), (e), (f).
 Within each set of four plots, the top right and bottom left images show (respectively) the observed and
 reconstructed beam image at the observation point; continuous lines in the top left and bottom right plots show
 the density projected onto (respectively) the horizontal and vertical axes, broken lines show Gaussian fits (used to
 determine the beam sizes, with values shown alongside the relevant plots).  Blue lines correspond to the observed image, and orange lines correspond to the
 reconstructed image. Note the different scales on the co-ordinate axes for 10\,pC and 100\,pC bunch charges.
\label{4dphasespacevalidationart}}
 \end{figure*}

Projections of the reconstructed 4D phase space distribution are shown in Fig.~\ref{4dphasespaceart} for 10\,pC and
100\,pC bunch charges.  Note that the scales on the axes for each image are given in normalised phase space
(units of mm/$\sqrt{\textrm{m}}$).  Validation  images for 10\,pC and 100\,pC bunch charges are shown in
Fig.~\ref{4dphasespacevalidationart} for three steps in the quadrupole scan.  The screen images are generally
reproduced from the co-ordinate space projection of the reconstructed phase space distribution with good
accuracy, supporting the validity  of the reconstructed 4D phase space distribution. The screen images with
100\,pC bunch charge show significanly more structure than those with 10\,pC bunch charge, though the
additional structure is not immediately apparent from the projections of the 4D phase space distribution at
the exit of the linac.  The richer beam structure observed with 100\,pC bunch charge is believed to be
associated with the properties of the photoinjector laser.
 
\begin{figure}
 \includegraphics[trim = 20pt 0pt 40pt 40pt, clip, width=\columnwidth,]{\figfile{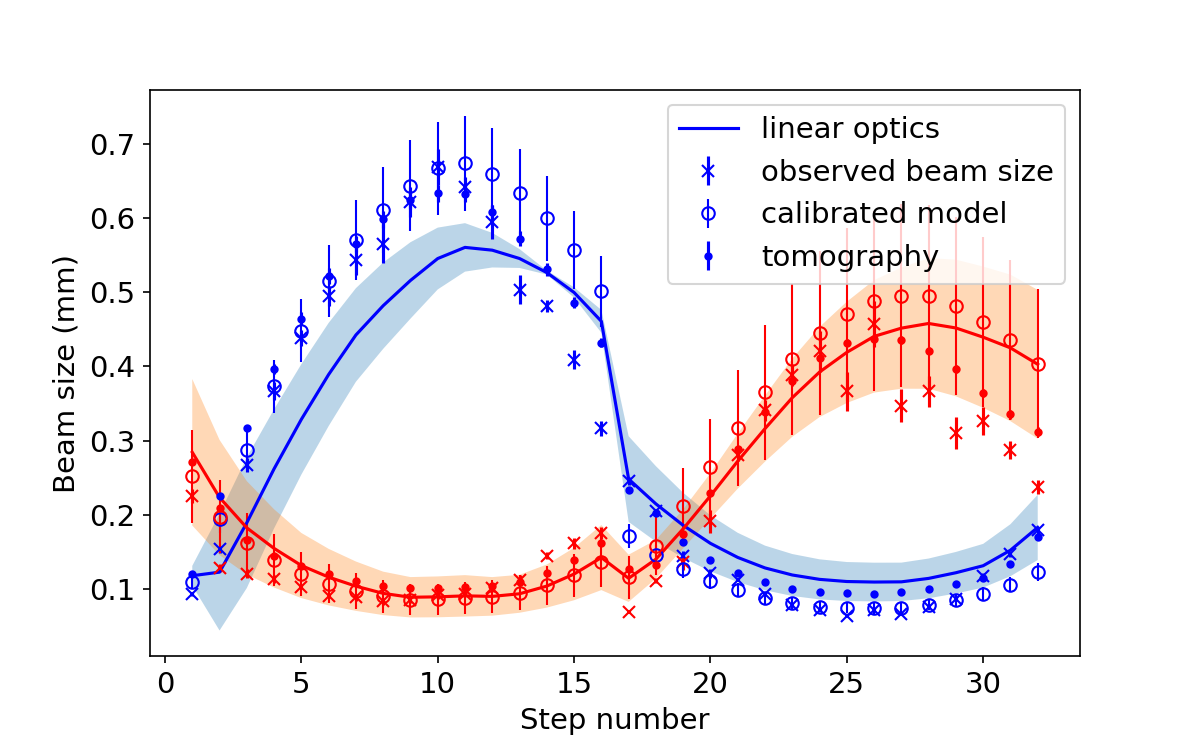}}
 \includegraphics[trim = 20pt 0pt 40pt 40pt, clip, width=\columnwidth,]{\figfile{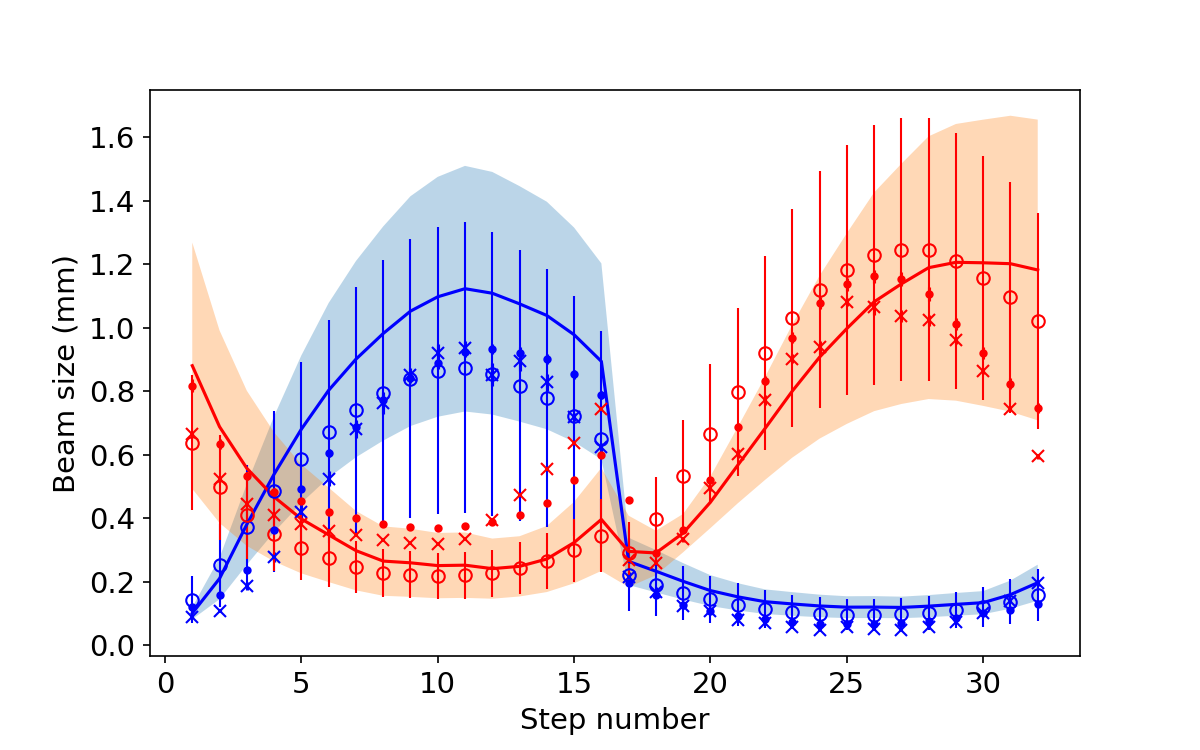}}
 \caption{Variation in horizontal (blue points and lines) and vertical (red points and lines) at the observation point,
 for 10\,pC bunch charge (top) and 100\,pC bunch charge (bottom).  Error bars on the observed beam sizes
 (marked as crosses) show the standard deviation of Gaussian fits to the ten beam images collected at the
 observation point for each step in the quadrupole scan.  Error bars on the beam sizes from the tomographic
 reconstruction (solid points) show the uncertainty in a Gaussian fit to the phase space density projected onto the
 horizontal or vertical axis.  Open circles show the beam sizes calculated by propagating the lattice functions for
 the calibrated model (Table~\ref{machineparameterstable}) from the reconstruction point to the observation
 point, and combining with the emittances calculated by a fit to the 4D phase space from ART tomography
 (Table~\ref{tablelatticefunctionsart}).  The line shows the beam sizes obtained by propagating the covariance
 matrix fitted to the 4D phase space distribution reconstructed by ART (Table~\ref{tablelatticefunctionsart}),
with shaded range showing the uncertainy arising from the uncertainties on the elements of the covariance matrix.
 \label{beamsizevariationart}}
 \end{figure}

Variations in the beam size at the observation point over the course of a quadrupole scan are shown in
Fig.~\ref{beamsizevariationart}.  The plots (upper plot for 10\,pC bunch charge, and lower plot for 100\,pC)
compare the beam sizes calculated in four different ways:
\begin{itemize}
\item The solid lines (labelled ``linear optics'') show the beam sizes (calculated at each point in the quadrupole scan)
found by calculating the covariance matrix describing the reconstructed 4D phase space distribution at the reconstruction
point, and then transporting the covariance matrix to the observation point.  The shaded bands indicate the
uncertainties on the beam sizes arising from the uncertainties on the elements of the covariance matrix.
\item Crosses (labelled ``observed beam size'' in Fig.~\ref{beamsizevariationart}) show the rms beam sizes
obtained from Gaussian fits to projections of the observed beam images onto the horizontal and vertical axes.
The error bars indicate the standard deviations of the rms beam sizes over the ten images collected at each
step (which dominate over uncertainties associated with the Gaussian fits).
\item The circular markers (labelled ``calibrated model'') show the beam sizes at each point in the quadrupole scan
expected from the lattice functions in the calibrated model, with emittances found from the reconstructed 4D phase space.
The error bars show the uncertainty arising from the uncertainty on the emittance (increased a factor of 10, to make
the error bars more clearly visible).
\item Points (dots, labelled ``tomography'') show the rms beam sizes obtained from Gaussian fits to projections
(onto the horizontal and vertical axes) of the reconstructed 4D phase space transported from the reconstruction
point to the observation point. The error bars in this case indicate the uncertainties in the fit.
\end{itemize}
Although there is qualitative agreement between the beam sizes in the calibrated model (using the optics functions
shown in Table~\ref{machineparameterstable}) and the observed beam sizes, there is better agreement with the
observed beam sizes in the case of linear transport of the covariance matrix calculated from the reconstructed
phase space distribution, and in the case of linear transport of the phase space distribution.

 For completeness, and for comparison of the results from tomographic analysis using ART and analysis
 using machine learning, the emittances and optics functions at the reconstruction point are given in
 Table~\ref{tablelatticefunctionsart}.  The values shown are calculated from the covariance matrices
 describing the reconstructed 4D phase space distributions, for 10\,pC and 100\,pC bunch charges.
 Note that the values given are for the normal mode emittances $\gamma\varepsilon_\mathrm{I}$,
 $\gamma\varepsilon_\mathrm{II}$ and optics functions, $B^\mathrm{I}$, $B^\mathrm{II}$
 \cite{wolski2006}.  In terms of these quantities, the covariance matrix is expressed:
 \begin{equation}
 \Sigma = \varepsilon_\mathrm{I} B^\mathrm{I} + \varepsilon_\mathrm{II} B^\mathrm{II},
 \end{equation}
 where the elements of the covariance matrix are the second-order moments of the beam distribution
 over all combinations of phase space variables:
 \begin{equation}
 \Sigma_{ij} = \langle x_i x_j \rangle,
 \end{equation}
 with $x_i = x, p_x, y, p_y$, for $i = 1, 2, 3, 4$, respectively.
The symmetric matrices $B^k$ can be written in terms of $2\times 2$ sub-matrices $\sigma^k_{uu}$
(with $u = x$ or $y$):
 \begin{equation}
 B^k = \left( \begin{array}{cc}
 \sigma^k_{xx} & \sigma^k_{xy} \\
 (\sigma^k_{xy})^\mathrm{T} & \sigma^k_{yy}
 \end{array} \right).
 \end{equation}
 In the absence of coupling:
 \begin{equation}
 \sigma^\mathrm{I}_{xx} = \left( \begin{array}{cc}
 \beta_x & -\alpha_x \\
 -\alpha_x & -\gamma_x
 \end{array} \right),
 \qquad
 \sigma^\mathrm{II}_{yy} = \left( \begin{array}{cc}
 \beta_y & -\alpha_y \\
 -\alpha_y & -\gamma_y
 \end{array} \right),
 \end{equation}
 and:
 \begin{equation}
 \sigma^\mathrm{I}_{yy} = \sigma^\mathrm{II}_{xx} = \sigma^\mathrm{I}_{xy} = \sigma^\mathrm{II}_{xy} = 0.
 \end{equation}

 \begin{table*}
 \caption{Emittances and lattice functions describing the 4D phase space distributions obtained by phase
 space tomography in CLARA.  The values given refer to the normal modes \cite{wolski2006}.  In the absence
of coupling, the elements of the matrices $\sigma^\mathrm{I}_{xx}$ and $\sigma^\mathrm{II}_{yy}$ are 
(respectively) $\beta_x$ and $\beta_y$ (top left elements) and $-\alpha_x$ and $-\alpha_y$ (top right elements);
and all other matrices are zero.\label{tablelatticefunctionsart}}
 \begin{tabular}{lcccc}
 \hline
  & \multicolumn{2}{c}{10\,pC} & \multicolumn{2}{c}{100\,pC} \\
  & algebraic reconstruction & machine learning & algebraic reconstruction & machine learning \\
\hline
\hline
  $\gamma\varepsilon_\mathrm{I}$ & $1.99\pm 0.04$\,{\textmu}m & $1.98\pm 0.03$\,{\textmu}m &
  $3.35\pm 0.44$\,{\textmu}m & $3.38\pm 0.16$\,{\textmu}m \\
  $\gamma\varepsilon_\mathrm{II}$ & $3.39\pm 0.19$\,{\textmu}m & $2.08\pm 0.06$\,{\textmu}m &
  $21.4\pm 1.7$\,{\textmu}m & $18.0\pm 0.3$\,{\textmu}m   \vspace*{6pt} \\
  $\sigma_{xx}^\mathrm{I}$ &
  $ \left( \begin{array}{cc}
   8.63\,\mathrm{m} & 1.06  \\
   1.06 & 0.245\,\mathrm{/m}
   \end{array} \right) $ 
   &
   $ \left( \begin{array}{cc}
   4.60\,\mathrm{m} & 0.260  \\
   0.260 & 0.200\,\mathrm{/m}
   \end{array} \right) $
   &
     $ \left( \begin{array}{cc}
    14.4\,\mathrm{m} & 0.782  \\
    0.782 & 0.112\,\mathrm{/m}
   \end{array} \right) $
   &
   $ \left( \begin{array}{cc}
    8.46\,\mathrm{m} & 0.389  \\
    0.389 & 0.136\,\mathrm{/m}
   \end{array} \right) $ \vspace*{3pt}
  \vspace*{3pt} \\
  $\sigma_{yy}^\mathrm{II}$ &
  $ \left( \begin{array}{cc}
   7.11\,\mathrm{m} & 2.14 \\
   2.14 & 0.786\,\mathrm{/m}
   \end{array} \right) $
   &
   $ \left( \begin{array}{cc}
   3.44\,\mathrm{m} & 1.06 \\
   1.06 & 0.572\,\mathrm{/m}
   \end{array} \right) $
   &
     $ \left( \begin{array}{cc}
  11.7\,\mathrm{m} & 3.67 \\
  3.67 & 1.24\,\mathrm{/m}
   \end{array} \right) $
  &
  $ \left( \begin{array}{cc}
  4.50\,\mathrm{m} & 1.62 \\
  1.62 & 0.804\,\mathrm{/m}
   \end{array} \right) $
  \vspace*{6pt} \\
  $\sigma_{xy}^\mathrm{I}$ &
  $ \left( \begin{array}{cc}
  -0.355\,\mathrm{m} & -0.0675 \\
  -0.0760 & -0.0221\,\mathrm{/m}
   \end{array} \right) $
   &
   $ \left( \begin{array}{cc}
  -0.933\,\mathrm{m} & -0.0708 \\
  -0.181 & -0.0897\,\mathrm{/m}
   \end{array} \right) $
   &
     $ \left( \begin{array}{cc}
  -1.02\,\mathrm{m} & -0.306 \\
  -0.100 & -0.0317\,\mathrm{/m}
   \end{array} \right) $
   &
   $ \left( \begin{array}{cc}
  -0.370\,\mathrm{m} & -0.148 \\
  -0.0306 & -0.0185\,\mathrm{/m}
   \end{array} \right) $
  \vspace*{3pt} \\
  $\sigma_{yy}^\mathrm{I}$ &
  $ \left( \begin{array}{cc}
  0.0238\,\mathrm{m} & 0.00667 \\
  0.00667 & 0.00218\,\mathrm{/m}
   \end{array} \right) $
   &
   $ \left( \begin{array}{cc}
  0.278\,\mathrm{m} & 0.0739 \\
  0.0739 & 0.0407\,\mathrm{/m}
   \end{array} \right) $
   &
     $ \left( \begin{array}{cc}
  0.101\,\mathrm{m} & 0.0314 \\
  0.0314 & 0.00978\,\mathrm{/m}
   \end{array} \right) $
  &
  $ \left( \begin{array}{cc}
  0.0177\,\mathrm{m} & 0.00781 \\
  0.00781 & 0.00375\,\mathrm{/m}
   \end{array} \right) $
  \vspace*{6pt} \\
  $\sigma_{xx}^\mathrm{II}$ &
  $ \left( \begin{array}{cc}
    0.0196\,\mathrm{m} & 0.00189 \\
    0.00189 & 0.000555\,\mathrm{/m}
   \end{array} \right) $
   &
   $ \left( \begin{array}{cc}
    0.334\,\mathrm{m} & 0.0242 \\
    0.0242 & 0.0194\,\mathrm{/m}
   \end{array} \right) $
   &
     $ \left( \begin{array}{cc}
   0.0376\,\mathrm{m} & -0.00173 \\
  -0.00173 & 0.000135\,\mathrm{/m}
   \end{array} \right) $
  &
  $ \left( \begin{array}{cc}
   0.0181\,\mathrm{m} & -0.000737 \\
  -0.000737 & 0.000330\,\mathrm{/m}
   \end{array} \right) $
  \vspace*{3pt} \\
  $\sigma_{xy}^\mathrm{II}$ &
  $ \left( \begin{array}{cc}
  0.245\,\mathrm{m} & 0.0342 \\
  0.0624 & 0.0197\,\mathrm{/m}
   \end{array} \right) $
   &
   $ \left( \begin{array}{cc}
  0.859\,\mathrm{m} & 0.0914 \\
  0.209 & 0.105\,\mathrm{/m}
   \end{array} \right) $
   &
     $ \left( \begin{array}{cc}
  0.148\,\mathrm{m} & -0.00876 \\
  0.0179 & 0.00863\,\mathrm{/m}
   \end{array} \right) $
  &
  $ \left( \begin{array}{cc}
  0.261\,\mathrm{m} & 0.0685 \\
  0.00419 & 0.0100\,\mathrm{/m}
   \end{array} \right) $
  \vspace*{3pt} \\
  \hline
 \end{tabular}
  \end{table*}

% ------------------------------------------------------------------------------

\section{Phase space tomography using machine learning\label{sec:ml}}

% ------------------------------------------------------------------------------

Although the results shown in Section~\ref{sec:tomography2dof} suggest that the algebraic
reconstruction technique can be of value in constructing the 4D transverse phase space distribution
of the beam in a machine such as CLARA, the method can have some limitations.  First, the
structures visible in the beam images at the observation point (especially at the higher bunch charge)
are not clearly evident in any of the projections shown of the 4D phase space distribution at the
reconstruction point.  The reasons for this are not well understood: it may simply be a result of the
relatively poor resolution with which the 4D phase space distribution is determined; or it may be that
the orientation of the distribution in phase space is such as to obscure the structure for the chosen
2D projections --- note that the structures seen at the observation point are only really evident for
particular steps in the quadrupole scan, i.e.~for some specific range of betatron phase advances.

A second limitation of the algebraic reconstruction technique is that it can take some time to process
the data to obtain the phase space distribution.  The demands in terms of processing time and
computational resources increase rapidly with increasing resolution of the reconstruction, and with
increasing dimensionality of the phase space.  For the results presented here, a phase space resolution of
39 pixels in each dimension of the 4D phase space is used: this limits the detail visible in the phase
space, but allows the reconstruction to be completed reasonably rapidly (within a few minutes) using
a standard PC.  Where a high resolution is required, or a rapid reconstruction would be of value (for
example, for several iterations of machine tuning) then more powerful computing resources
may be needed if algebraic reconstruction, or a similar tomography technique, is to be used.
There is also interest in extending tomography from four to five or six dimensions
\cite{edelen2020, scheinker2022}: this can be of particular value in short-wavelength free electron
lasers, for example, where understanding the transverse beam profile and energy spread as a function
of longitudinal position in the bunch can be of significant importance.

Approaches based on machine learning may offer ways to address some of the issues associated
with conventional tomography techniques for reconstruction of the beam phase space in four (or more)
dimensions.  The method presented here, which we apply to the two transverse degrees of freedom, uses
a pre-trained neural network, to which the beam images at the observation point are provided, in
compressed form, as input; the output from the neural network consists of the 4D phase space
distribution, again in compressed form.  In principle, using a neural network in this way allows a rapid
(almost immediate) reconstruction of the 4D phase space distribution once the beam images are provided.
The computing resources needed for carrying out the reconstruction can also be much more modest than those
needed for algebraic reconstruction tomography.  If images in uncompressed form are used, the input and
output data sets can still be of significant size, but use of machine learning enables image compression
techniques to be applied, reducing the size of input and output data sets.  In principle, a neural network
can be trained on images and phase space distributions represented in some chosen compressed
form, for example as discrete cosine transforms (DCTs)
\cite{ahmednatarajanrao1974,chenpratt1984,raoyip1990}.  Image compression would be difficult to apply
in the case of conventional tomography methods, which usually rely on a relationship between the
sinogram and the object to be reconstructed that is intrinsically expressed in regular co-ordinate space.
Neural networks offer much greater flexibility, and do not require a specific representation
of the input or output data.

In using a neural network to perform tomographic reconstruction, an issue does arise with the need
to train the network.  Training must necessarily be based on simulated data, which would
ideally include features characteristic of the beam; but at least in cases where the beam shows some
detailed structure, the relevant features may not be known at the time of generating the training data.
In the current study, we simply take the approach of generating random phase spaces consisting of a
number of superposed 4D Gaussian distributions, with the component distributions in each generated
phase space varying randomly in position, shape and intensity. Given the shape of the phase space
distribution in CLARA suggested by tomography using ART, the phase space distributions constructed
in this way may not provide ideal training data; however, it is interesting to consider the ability of a neural
network to reconstruct phase space distributions presenting features significantly different from those
present in the training data.  If the techniques described here are to be of value in a reasonably wide
range of situations, then they should be able to reproduce phase space distributions with features
significantly different from those in the training data.

% ------------------------------------------------------------------------------

\subsection{Implementation of machine learning method\label{sec:mlimplementation}}

% ------------------------------------------------------------------------------

Before presenting the results of tomography using machine learning, we discuss some further
details of how the technique was implemented. 

For preparation of training data, phase space distributions were generated as mentioned above, 
by superposing 4D Gaussian distributions with random variations in position, shape and intensity.
The distributions are constructed in normalised phase space; the sinograms are then obtained by
transforming the distribution using phase space rotations (corresponding to the steps in a quadrupole
scan), and then projecting the distribution onto the (normalised) $x$--$y$ plane at each step in the
quadrupole scan.  Note that we used phase advances corresponding to those in the calibrated model,
shown in Fig.~\ref{opticsfunctionsdesignandcalibrated} (bottom right).  For consistency, it is important
that the phase advances should match those resulting from the quadrupole strengths applied in the
quadrupole scan, given the lattice functions used for normalising the phase space.  It should be
emphasised, however, that the chosen lattice functions do not need to match those describing the
actual beam distribution (which in general, is not known in advance).

\begin{figure}
 \begin{center}
 \includegraphics[trim = 65pt 80pt 65pt 100pt, clip, width=\columnwidth,]{\figfile{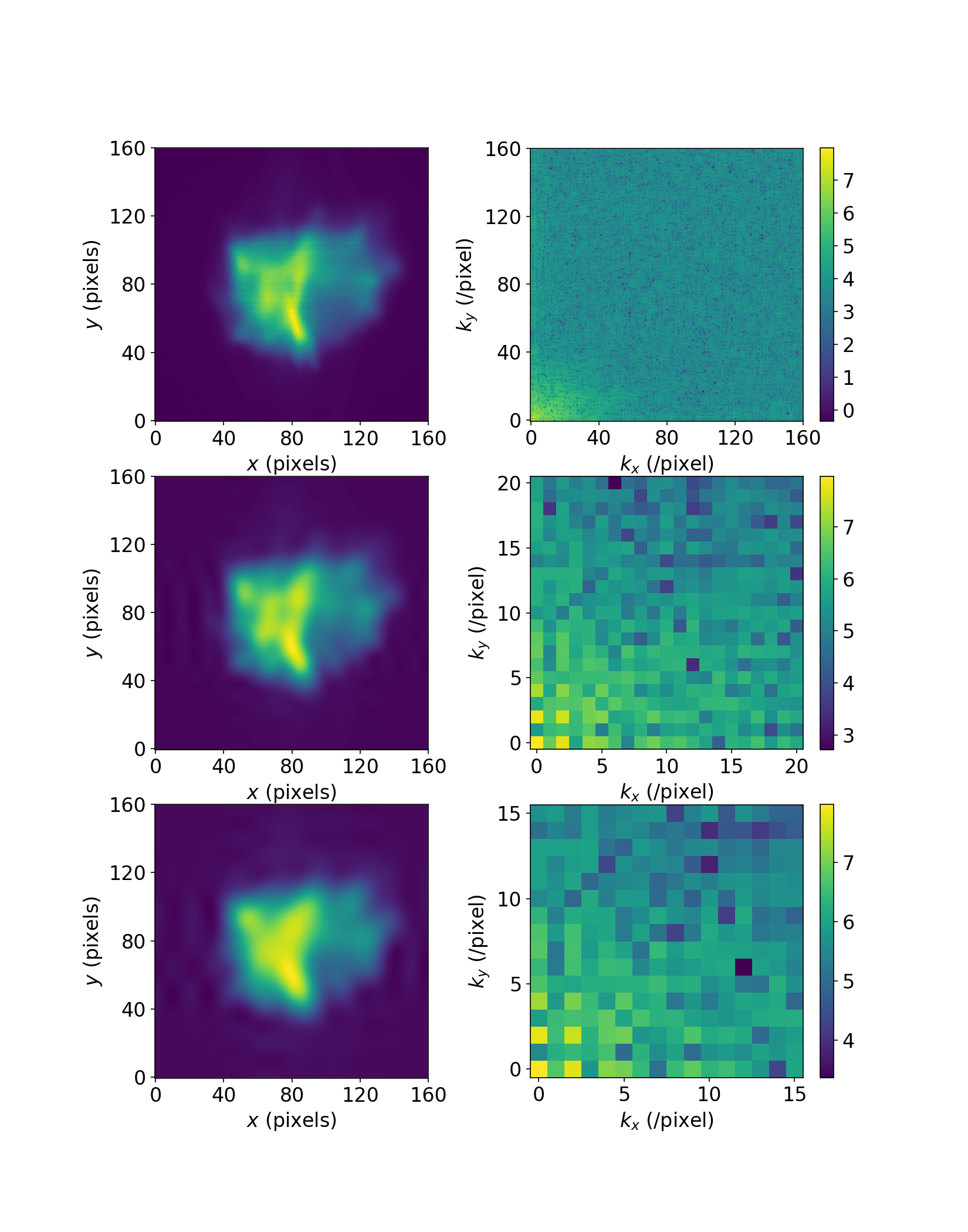}}
 \end{center}
 \caption{Image compression using truncation of the DCT.  In each pair, the left-hand image
 shows the beam image in co-ordinate space reconstructed from the DCT shown in the right-hand image.
 Images are reconstructed at the resolution of the original image by padding the truncated DCT with
 zeros as necessary.   Top: full resolution, 161$\times$161 pixels.  Middle: DCT truncated
 to 21$\times$21.  Bottom: DCT truncated to 16$\times$16.  Beam images are from a quadrupole scan with
 100\,pC bunch charge.  The colour scale shows the logarithm (to base 10) of the absolute value of the
 DCT component.
 \label{imagecompressiondct}}
 \end{figure}

Having obtained the sinograms for the simulated 4D phase space distributions, we compress both
the phase space distributions and the sinograms using discrete cosine transforms (DCTs).  There are several
types of DCT: we use a Type II DCT, which is the default in many standard scientific computing
packages.  In the case of a 2D $M\times N$ array, a Type II DCT is defined by:
\begin{equation}
y_{jk} = \sum_{m=0}^{M-1}\sum_{n=0}^{N-1} x_{mn} \cos\!\left( \pi j \frac{2m+1}{2M} \right)\cos\!\left( \pi k \frac{2n+1}{2N} \right),
\label{dctdef}
\end{equation}
where the values $x_{mn}$ are the components of the initial array, and $y_{jk}$ (for $j = 0 \ldots M-1$,
$k = 0 \ldots N-1$)
are the components of the transformed array.  Compression is achieved by truncating the transformed
array at some point, either defined in terms of the magnitudes of the components (which should all be
below some specified threshold beyond the truncation point) or simply in terms of a fixed limit on the
size of the transformed array.  The inverse of the Type II DCT of an $M\times N$ array is given by:
\begin{eqnarray}
 & & x_{mn} = \frac{1}{MN} \times \nonumber \\
 & &  \sum_{j=0}^{M-1}\sum_{k=0}^{N-1} \alpha_{jk} y_{jk}
\cos\!\left( \pi j \frac{2m+1}{2M} \right)\cos\!\left( \pi k \frac{2n+1}{2N} \right),  \nonumber \\
 & & 
\label{invdct}
\end{eqnarray}
where:
\begin{equation}
\alpha_{jk} = \left\{ \begin{array}{ll}
1 & \textrm{ if } j = k = 0, \\
2 & \textrm{ if } j = 0, k\neq 0, \textrm{ or } j\neq 0, k = 0, \\
4 & \textrm{ if } j\neq 0, k\neq 0.
\end{array} \right.
\end{equation}
%$\alpha_{jk} = 1$ if $j = k = 0$, $\alpha_{jk} = 2$ if $j=0$, $k\neq 0$ or $j\neq 0$, $k=0$,
%and $\alpha_{jk} = 4$ if $j\neq 0$ and $k\neq 0$.
The expressions in (\ref{dctdef}) and (\ref{invdct}) can be extended to higher-dimensional arrays by including an additional
summation for each additional index, and making the appropriate modification to the numerical factors in
(\ref{invdct}).
Truncating the transformed array corresponds to reducing the upper limits on the summations in the inverse
transformation (\ref{invdct}); in this case, the array $x_{mn}$ is reconstructed with approximated values for
its elements, but the number of elements in the array remains the same. In the case of an image, the
effect of truncating the DCT is to lose some of the fine detail.  Figure \ref{imagecompressiondct} illustrates
image compression using DCTs truncated to different sizes, using (as an example) a beam image collected
during the quadrupole scan with 100\,pC bunch charge.  The original image has resolution (in pixels)
$M\times N = 161\times 161$.  Truncating the DCT to $21\times 21$ results in some loss of clarity, but the main features
and some details can still be clearly seen. Truncation to $16\times 16$ results in more significant loss of detail. 

\begin{figure*}
 \begin{center}
 \includegraphics[trim = 1pt 1pt 1pt 1pt, clip, width=2\columnwidth,]{\figfile{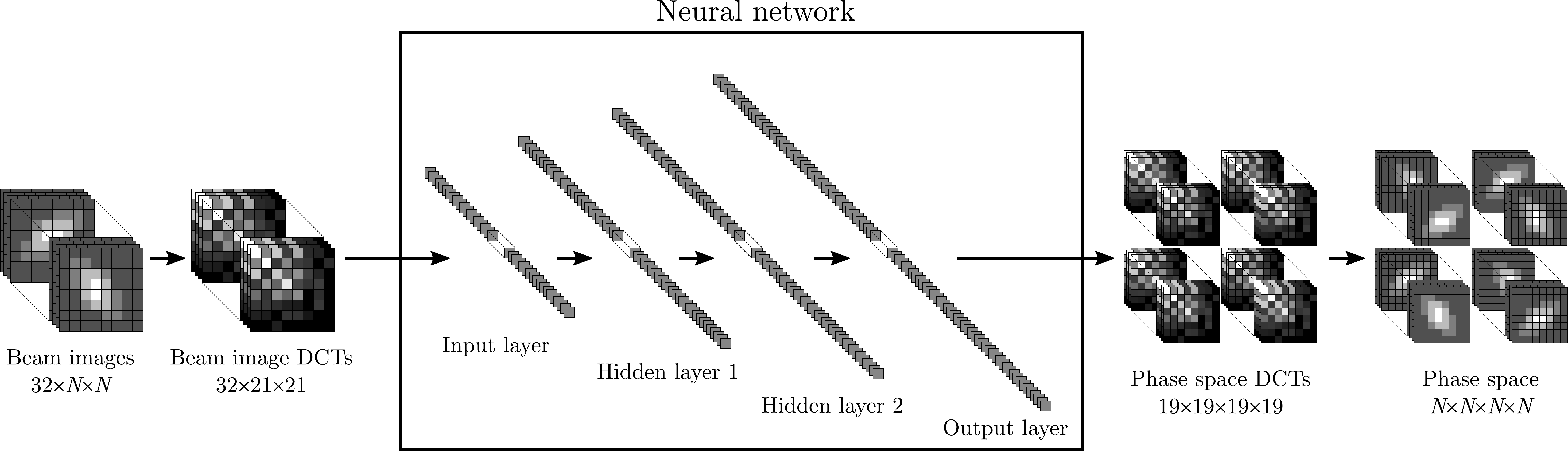}}
 \end{center}
 \caption{Schematic showing the steps in the phase space tomography using image compression and
machine learning.  The beam images collected during a quadrupole scan are compressed by applying
a discrete cosine transform (DCT) to each image.  The transformed (and compressed) images are provided
as input to a neural network, consisting of an input layer, two hidden (dense) layers, and an output layer.
The output of the neural network is a DCT of the 4D phase space distribution of the
beam: applying an inverse DCT allows reconstruction of the phase space density at any desired resolution.
 \label{dataprocessingschematic}}
 \end{figure*}

The training data for the neural network consists of some number of pairs of the DCTs of the
sinograms (input) and corresponding phase space distributions (output).  The neural network
itself is implemented in Keras \cite{keras}.  We use a rather straightforward architecture.  Apart from
the input and output layers, there are two hidden layers, defined as dense layers in Keras.  To limit
overtraining, each dense layer is followed by a dropout layer.  We use a resolution of 19 points on each
axis for the DCT of the 4D phase space (i.e.~$19^4$ voxels in total), and a resolution of $21\times 21$
for the DCT of each 2D projection in the set of ``images'' forming the sinogram.  In practice, these
resolutions capture sufficient numbers of DCT modes to allow representation of the screen images and
the 4D phase space with good resolution.  Note that the size of the data for the 4D phase space using
machine learning ($19^4$) is substantially smaller than the size used for the ART tomography reported in
Section~\ref{sec:tomography2dof} ($39^4$).  We have found that for the data collected in CLARA,
increasing the numbers of DCT modes, either in the input sinograms or the reconstructed phase space,
does not improve the quality of the results as judged by a comparison between the projections of the
phase space at the observation point, and the original beam images (as shown, for example, in 
Fig.~\ref{4dphasespacevalidationml}). In constructing the sinogram, we
use phase space rotations corresponding to the phase advances in the calibrated model (see
Fig.~\ref{opticsfunctionsdesignandcalibrated}, bottom right plot), i.e.~with 32 steps in the quadrupole
scan.  With these parameters, the neural network has an input layer with $32\times 21^2$ nodes,
and an output layer with $19^4$ nodes.  We use 1500 and 3000 nodes for the first and
second hidden (dense) layers, respectively, with a dropout layer specified to set 20\% of inputs
(selected randomly) to zero for each dense layer during training.  The tomography process using
image compression and machine learning is illustrated schematically in Fig.~\ref{dataprocessingschematic}.

A total of 3000 sets of 4D phase space distributions and sinograms were generated as training data;
100 sets were reserved as validation sets for testing the performance of the trained network, and were
not used in the training process itself.  Training was carried out using the Adam optimization algorithm
\cite{adamoptimization}.  Training takes several minutes on a standard laptop PC.  The training time is
comparable to the time taken to process a single data set using ART; however, training only needs to be
performed once, to produce a neural network that can (in principle) be applied to any data set collected
in a quadrupole scan using given quadrupole strengths.  The ART analysis would need to be performed
separately for each data set.

\begin{figure*}
 \begin{center}
 \includegraphics[trim = 140pt 40pt 100pt 10pt, clip, width=2\columnwidth,]{\figfile{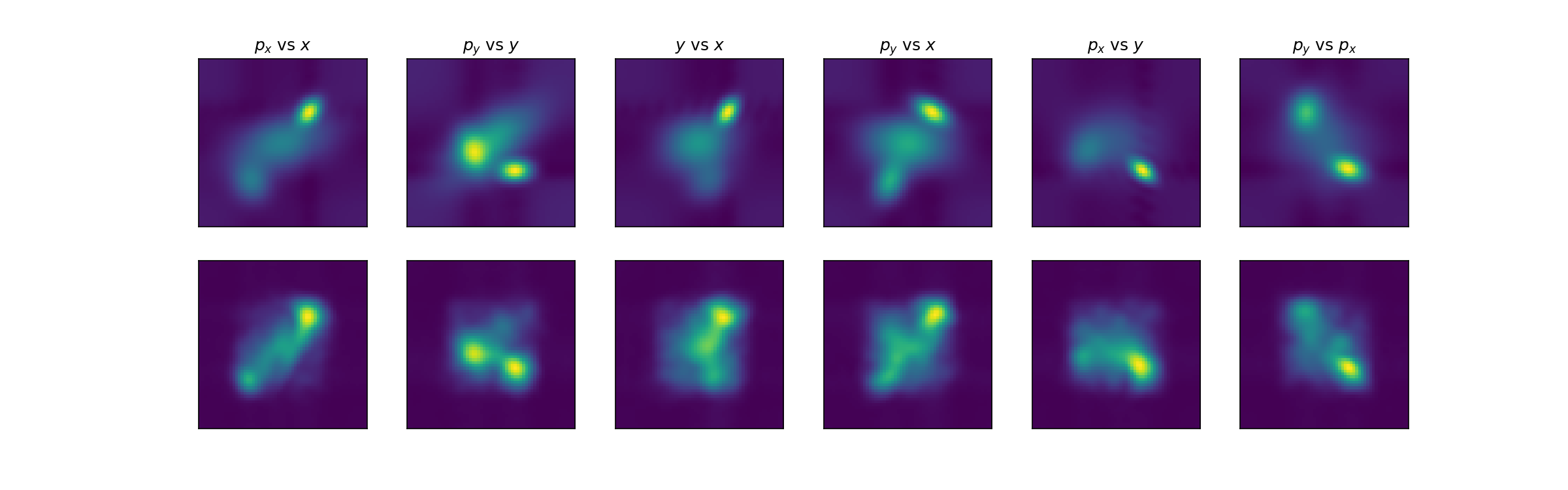}}
 Example 1 \\
 \includegraphics[trim = 140pt 40pt 100pt 10pt, clip, width=2\columnwidth,]{\figfile{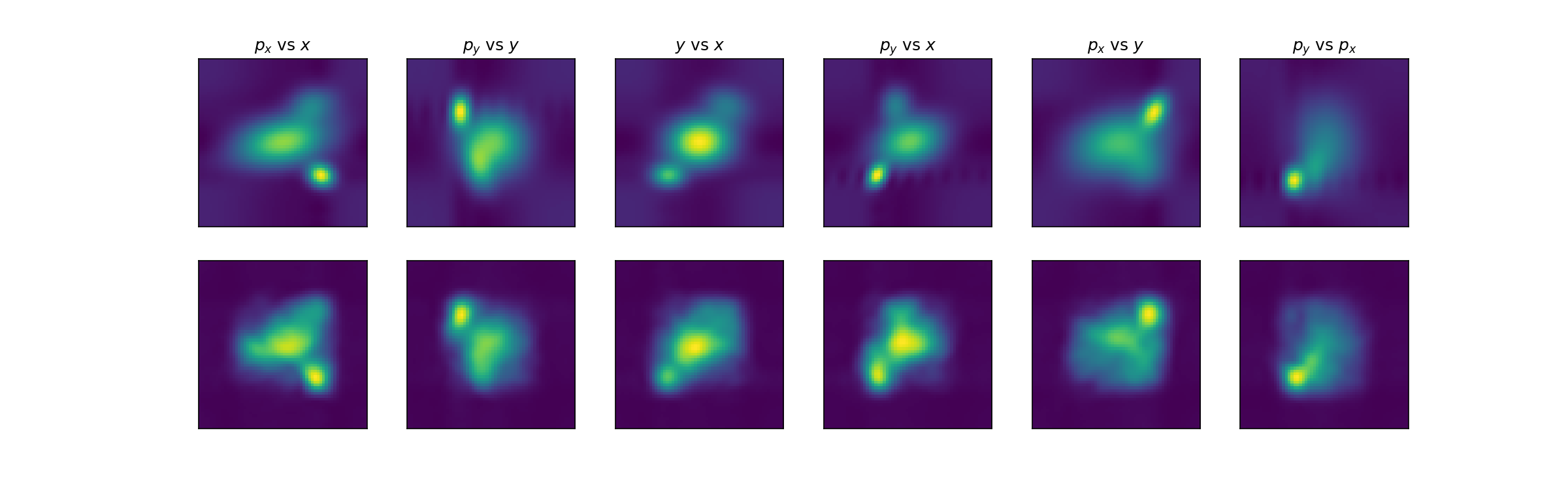}}
 Example 2
 \end{center}
 \caption{Examples of reconstruction of charge density in 4D phase space using a neural network.
Two different cases are shown.  In each example, plots in the top row show different projections of
the original phase space distribution from which the sinograms (projections onto $x$--$y$ co-ordinate
space following phase space rotations corresponding to steps in a quadrupole scan) are constructed.
Plots in the bottom row show the corresponding projections, from the 4D phase space reconstructed
from the sinograms using the neural network.
 \label{mltrainingexamples}}
 \end{figure*}

Two examples illustrating results from the trained network are shown in
Fig.~\ref{mltrainingexamples}.  The examples are selected at random from the validation data sets.
Each row of images in the figure shows a different projection of a
4D phase space: in each example, the top row shows the projections from the original phase space, 
and the bottom row shows the projections from the phase space reconstructed by the neural network
when provided with the (DCTs of the) corresponding sinograms.  While there are clearly some differences
between the original and the reconstructed phase spaces, the reconstruction is sufficiently similar to
the original to provide a useful practical indication of the beam distribution in phase space.

\begin{figure}
 \begin{center}
 \includegraphics[trim = 1pt 1pt 1pt 1pt, clip, width=\columnwidth,]{\figfile{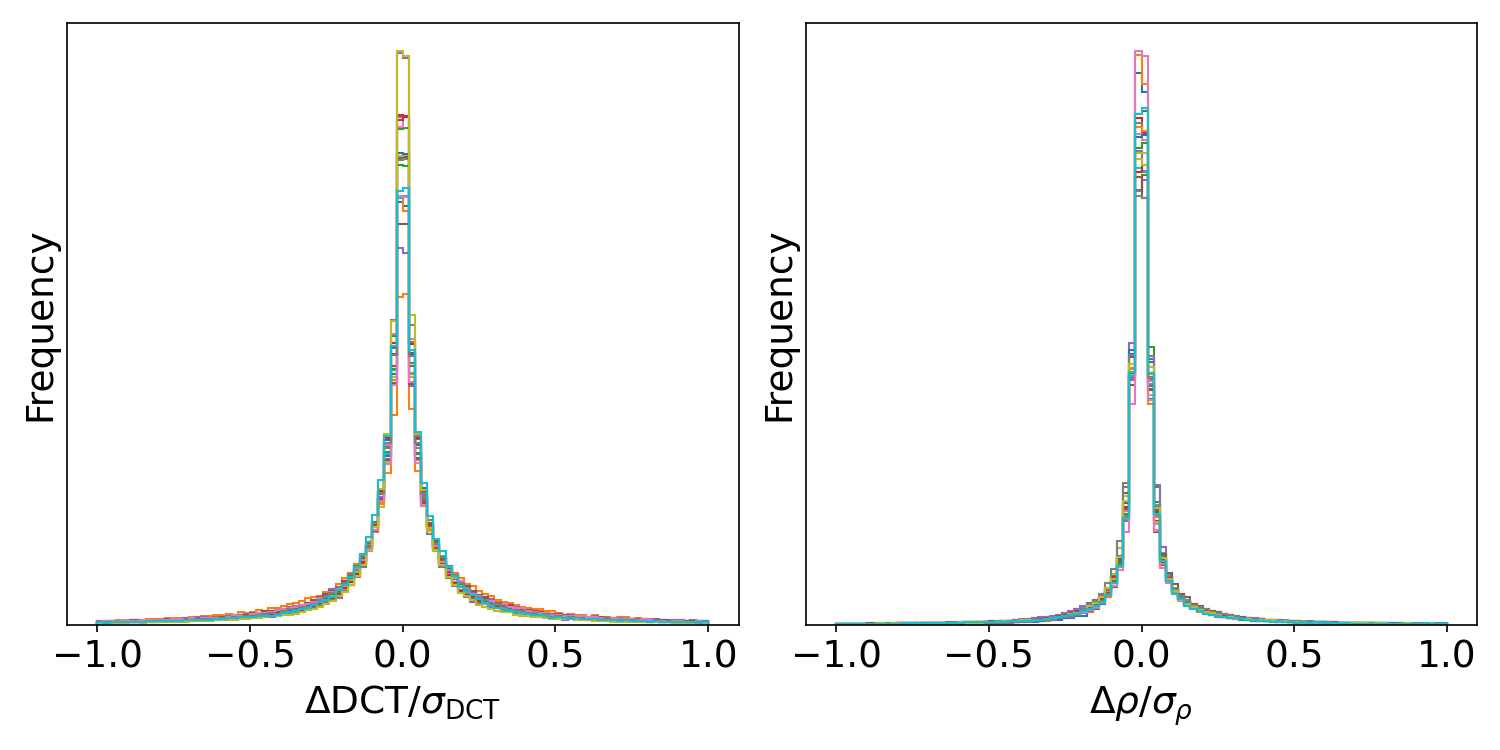}}
 \end{center}
 \caption{Residuals of the fit to the phase space density using the trained neural network.
 Left: histograms showing the frequency of different values of $\Delta$DCT/$\sigma_\mathrm{DCT}$,
 where $\Delta$DCT are the differences between the known DCT values (in one case from the test data)
 and the values found by the neural network from the corresponding sinograms, and $\sigma_\mathrm{DCT}$
 is the standard deviation of the DCT values.  Right: histograms showing the same residual analysis, but
 using the phase space densities, rather than the DCT of the densities.  20 cases are superposed in each
 plot: there is little variation between the cases in the distributions of residuals.
 \label{mlresiduals}}
 \end{figure}

To characterise further the reliability of the machine learning reconstruction of the phase space, we
calculate the residuals between the original phase space density in the test data and the phase space
density found from the sinograms using the trained neural network.  The residuals are shown in
Fig.~\ref{mlresiduals}, as histograms of $\Delta\mathrm{DCT}/\sigma_\mathrm{DCT}$ and
$\Delta\rho / \sigma_\rho$.  Here, $\Delta\mathrm{DCT}$ is the difference between a particular
DCT coefficient predicted by the neural network, and the corresponding DCT coefficient in the phase
space distribution used to generate the sinogram data provided as input to the network.
$\sigma_\mathrm{DCT}$ is the standard deviation of the DCT coefficients.  $\Delta\rho$ is the
difference in the phase space density (at a particular element of 4D phase space) between the
original distribution and the distribution found by the neural network, after performing an inverse
DCT of the network output; and $\sigma_\rho$ is the standard deviation of the phase space density.
Figure \ref{mlresiduals} shows histograms of these quantities for 20 cases from the validation data sets.
Typically, between 75\% and 80\% of phase space density values from the neural network are within
0.1\,$\sigma_\rho$ of the true phase space density.

\subsection{Experimental results from tomography using machine learning}

\begin{figure*}
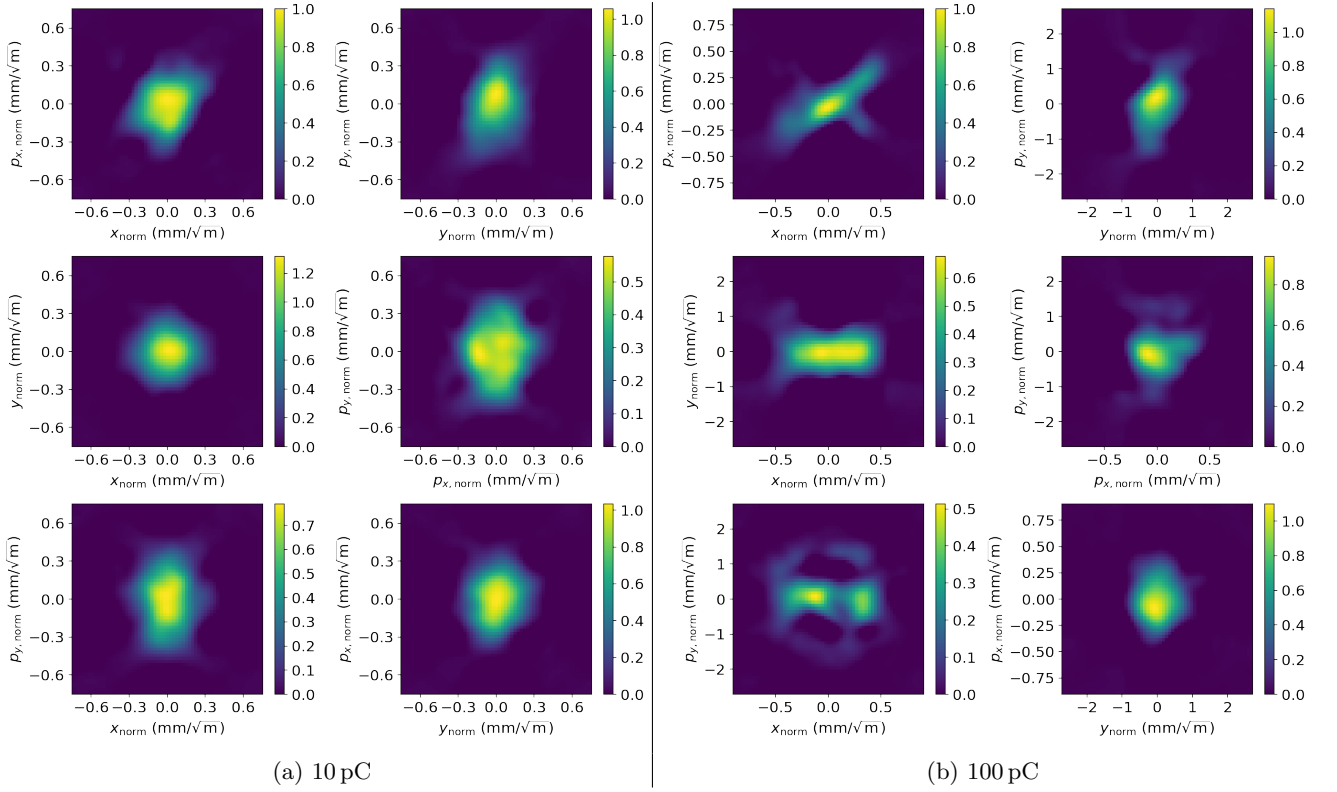

 \begin{tabular}{c|c}
 \includegraphics[trim = 0pt 35pt 65pt 80pt, clip, width=\columnwidth,]{\figfile{PS4DPhaseSpaceProjections-ML-10pC}} &
 \includegraphics[trim = 0pt 35pt 65pt 80pt, clip, width=\columnwidth,]{\figfile{PS4DPhaseSpaceProjections-ML-100pC}} \\
  (a) 10\,pC &
  (b) 100\,pC
 \end{tabular}
 \caption{Projections of the 4D phase space distribution of the beam in CLARA at the  exit of the linac, for
 (a) 10\,pC bunch charge and (b) 100\,pC bunch charge, found from machine learning.
 \label{4dphasespaceml}}
 \end{figure*}

The trained neural network was applied to analysis of the quadrupole scan data collected on CLARA,
described in Section~\ref{sec:tomography2dof}.  The screen images from each step of the quadrupole
scan were prepared in the same way as for the ART analysis, by cropping, and then scaling to transform to
normalised phase space.  The images were then compressed by constructing the DCTs, which were
truncated to 21 modes on each axis.  The DCTs were provided as input to the trained
neural network, which provided the DCT of the 4D phase space distribution, with resolution 19 modes
along each axis.  Projections from the reconstructed 4D phase space distribution for 10\,pC and 100\,pC
bunch charges are shown in Fig.~\ref{4dphasespaceml}.

\begin{figure*}
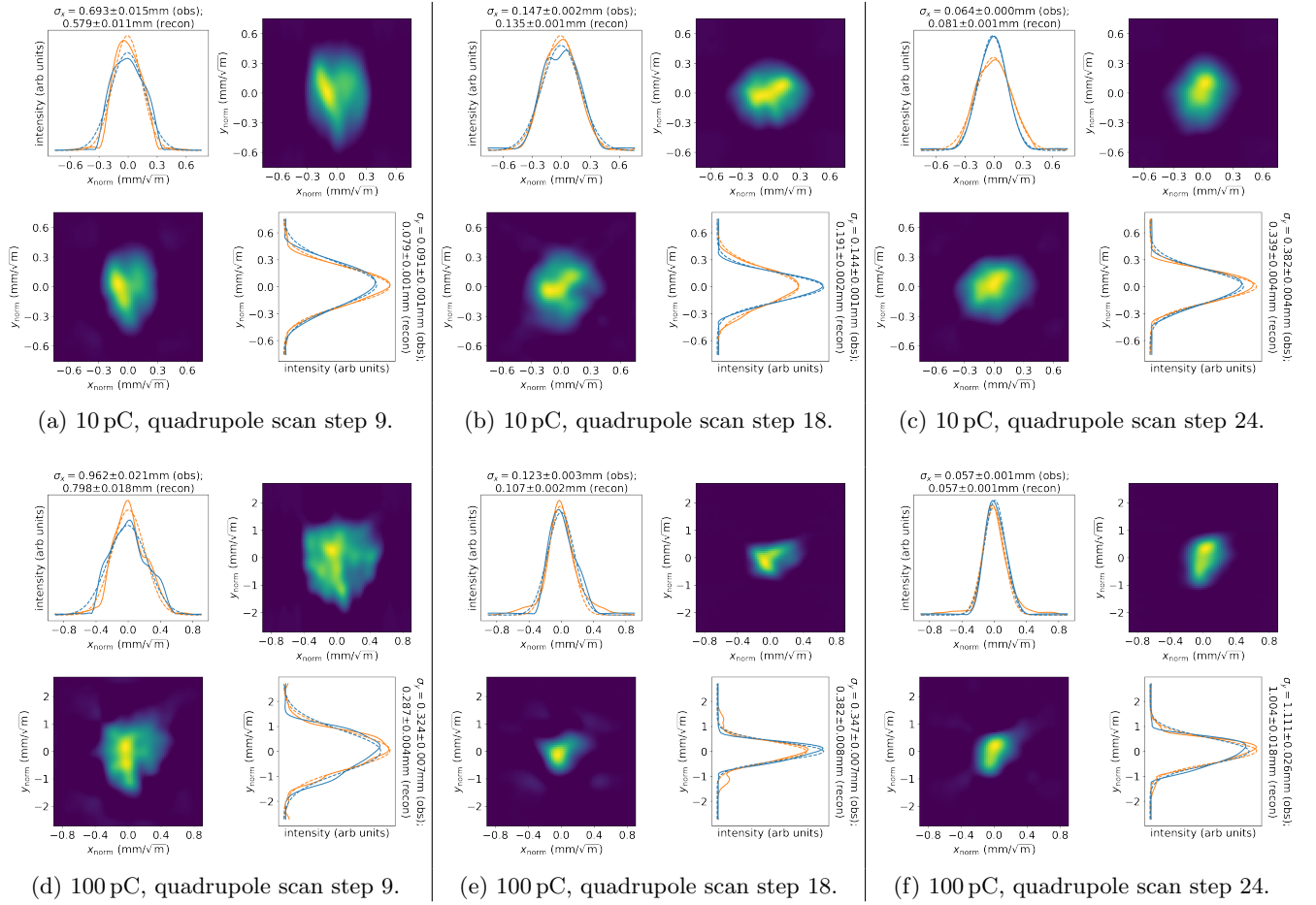

 \begin{tabular}{c|c|c}
 \includegraphics[trim = 25pt 35pt 60pt 60pt, clip, width=0.67\columnwidth,]{\figfile{PS4DTomographySinograms-ML-Resn19-09-10pC}} &
 \includegraphics[trim = 25pt 35pt 60pt 60pt, clip, width=0.67\columnwidth,]{\figfile{PS4DTomographySinograms-ML-Resn19-18-10pC}} &
 \includegraphics[trim = 25pt 35pt 60pt 60pt, clip, width=0.67\columnwidth,]{\figfile{PS4DTomographySinograms-ML-Resn19-24-10pC}} \\
 (a) 10\,pC, quadrupole scan step 9. &
 (b) 10\,pC, quadrupole scan step 18. &
 (c) 10\,pC, quadrupole scan step 24. \\[12pt]
 \includegraphics[trim = 25pt 35pt 60pt 60pt, clip, width=0.67\columnwidth,]{\figfile{PS4DTomographySinograms-ML-Resn19-09-100pC}} &
 \includegraphics[trim = 25pt 35pt 60pt 60pt, clip, width=0.67\columnwidth,]{\figfile{PS4DTomographySinograms-ML-Resn19-18-100pC}} &
 \includegraphics[trim = 25pt 35pt 60pt 60pt, clip, width=0.67\columnwidth,]{\figfile{PS4DTomographySinograms-ML-Resn19-24-100pC}} \\
 (d) 100\,pC, quadrupole scan step 9. &
 (e) 100\,pC, quadrupole scan step 18. &
 (f) 100\,pC, quadrupole scan step 24.
 \end{tabular}
 \caption{Validation images for 10\,pC bunch charge, found from machine learning, from
 three steps in the quadrupole scan.  Within each set of four plots, the top
 right and bottom left images show (respectively) the observed and reconstructed beam image at the observation
 point; continuous lines in the top left and bottom right plots show the density projected onto (respectively) the
 horizontal and vertical axes, broken lines show Gaussian fits (used to
 determine the beam sizes, with values shown alongside the relevant plots).  Blue lines
 correspond to the observed image, and orange lines correspond to the reconstructed image. 
\label{4dphasespacevalidationml}}
 \end{figure*}

In Section~\ref{sec:tomography2dof}, we validated the ART reconstruction of the 4D phase space
distribution by comparing the projection of the distribution onto $x$--$y$ co-ordinate space at the
observation point with the observed beam images at different steps of the quadrupole scan.  We can
make similar comparisons to validate the 4D phase space distribution reconstructed using the neural
network: some examples (for the same steps as shown in Fig.~\ref{4dphasespacevalidationart}) are
shown in Fig.~\ref{4dphasespacevalidationml}.  Once again, we see generally good agreement between
the projection of the 4D phase space distribution and the observed images, in both the 10\,pC and the
100\,pC cases.  Comparing with projections from the phase space reconstructed using ART in
Fig.~\ref{4dphasespacevalidationart}, the machine learning projections do not all have the same
clarity, in terms of the finer details in some of the images.  It should be remembered, however, that
the ART tomography uses beam images with resolution 39$\times$39 pixels, to reconstruct the 4D
phase space distribution with a resolution of 39 pixels on each axis.  The machine learning technique
uses beam images and 4D phase space in a compressed form: the beam images are represented by
21 DCT modes on each axis, and the phase space is represented by 19 DCT modes on each axis.
Although this is sufficient to capture a significant amount of detail, the truncation of the DCTs means
that the compression is not lossless.  Given the compression ratio, the machine learning method
retains a reasonable level of detail in the phase space distribution. 

\begin{figure}
 \includegraphics[trim = 20pt 0pt 40pt 40pt, clip, width=\columnwidth,]{\figfile{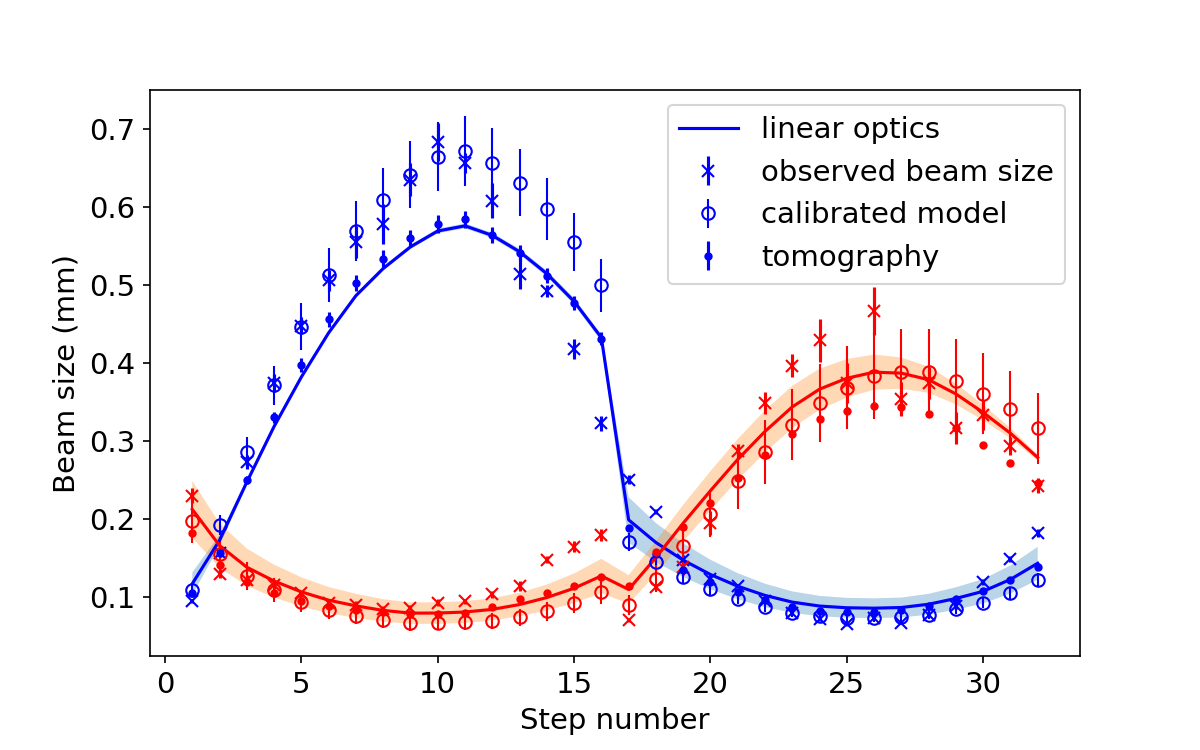}}
 \includegraphics[trim = 20pt 0pt 40pt 40pt, clip, width=\columnwidth,]{\figfile{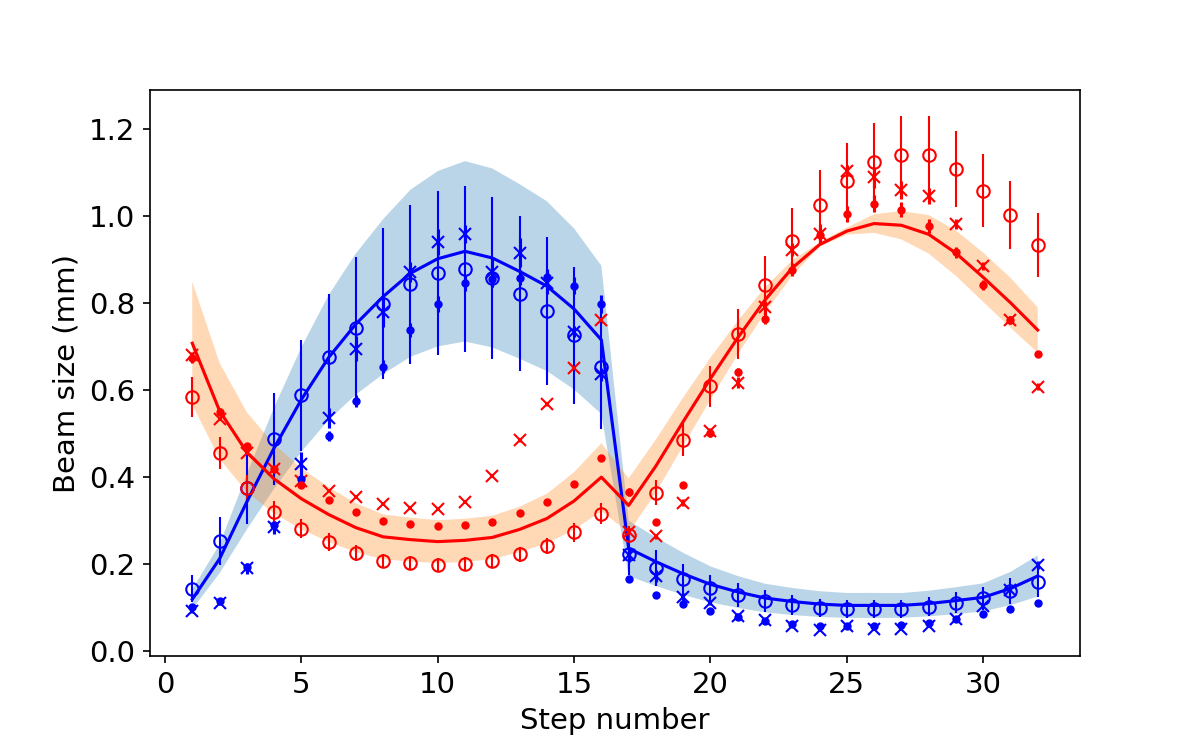}}
 \caption{Variation in horizontal (blue points and lines) and vertical (red points and lines) at the observation point,
 for 10\,pC bunch charge (top) and 100\,pC bunch charge (bottom).  Error bars on the observed beam sizes
 (marked as crosses) show the standard deviation of Gaussian fits to the ten beam images collected at the
 observation point for each step in the quadrupole scan.  Error bars on the beam sizes from the tomographic
 reconstruction (solid points) show the uncertainty in a Gaussian fit to the phase space density projected onto the
 horizontal or vertical axis.  Open circles show the beam sizes calculated by propagating the lattice functions for
 the calibrated model (Table~\ref{machineparameterstable}) from the reconstruction point to the observation
 point, and combining with the emittances calculated by a fit to the 4D phase space from machine learning
 (Table~\ref{tablelatticefunctionsart}).  The line shows the beam sizes obtained by propagating the covariance
 matrix fitted to the 4D phase space distribution reconstructed by machine learning
 (Table~\ref{tablelatticefunctionsart}).
 \label{beamsizevariationml}}
 \end{figure}

Comparisons between the observed and reconstructed beam sizes are shown in
Fig.~\ref{beamsizevariationml}: the results here can be compared with those in
Fig.~\ref{beamsizevariationart}, which shows the beam sizes reconstructed using ART.  While there
are some differences in detail in the quality of the match between the beam sizes expected from the
reconstructed phase space and the beam sizes observed during the quadrupole scan, both the
ART and the machine learning techniques show similar performance in describing the beam behaviour.
 
% ------------------------------------------------------------------------------

\section{Conclusions and possible further developments\label{sec:conclusions}}

% ------------------------------------------------------------------------------

The machine learning technique we have described in this paper uses relatively simple methods
for reconstructing the 4D phase space.  Nevertheless, this approach appears capable of producing
useful results, as shown by the comparison between projections onto $x$--$y$ co-ordinate space at
the observation point for different quadrupole strengths, and beam images collected over the
course of a quadrupole scan.  Values obtained for parameters describing the distribution (emittances
and lattice functions) are consistent with those obtained using a conventional tomography technique.
Data collection and analysis were planned using a design model of the machine; despite significant
differences between the design model and the actual machine conditions during collection
of experimental data, results from both the ART and the machine learning techniques provide useful
information on the beam properties in CLARA.

Use of image compression (in the present case, using discrete cosine transforms) allows reduction
of the size of the data sets that need to be processed, in particular, for representing the 4D phase space.
Machine learning allows direct tomographic analysis of compressed beam images and phase space
representations, without the additional complications or difficulties that would be encountered in attempting
to apply conventional tomography techniques to compressed images.

Inspection of projections of the 4D phase space onto various planes (in particular, comparison of
Fig.~\ref{4dphasespaceart} with Fig.~\ref{4dphasespaceml})
suggests that the machine learning technique is capable of producing a representation of the 4D phase
space distribution that appears clearer than that obtained by the conventional tomography algorithm.
On the other hand, the beam images obtained by projecting the 4D phase space distribution onto
co-ordinate space at the observation point
have slightly higher fidelity in the case of the conventional tomography technique.  Nevertheless, the
consistency in the results of the two methods, and the fact that the neural network produces a
4D phase space distribution immediately the quadrupole scan images are available (compared to a
potentially lengthy computation time required by the conventional tomography technique) suggests that
the machine learning approach could have some practical value.

There are a number of ways in which the machine learning approach could be further developed.
With an improved understanding of the operational conditions of CLARA, some optimisation would be
possible in terms of the quadrupole strengths (and number of steps) used in the quadrupole scan.
More sophisticated neural network architectures, or use of more sophisticated machine learning tools
generally, could lead to a better reconstruction of the 4D phase space distribution from a given set of
sinograms.  There may be some benefits in further increasing the number of sets of training data.
An indication of the quality to be expected in the reconstruction can be obtained using simulated data,
for example by calculating the residuals as shown in Fig.~\ref{mlresiduals}.  Although the phase space
distributions in the training data we used for the neural network had very different features from the
phase space distribution in the real machine, the trained network was still capable of reconstructing
a phase space distribution that provided a good description of beam behaviour.  It is possible, however,
that using training data more closely resembling the real beam (once some initial characterisation of the
beam has been obtained) could lead to better results.

Discrete cosine transforms may not be the optimal way to represent images and phase space distributions
in compressed form for the application described here.  A DCT essentially represents a multidimensional
array as a set of orthogonal modes, with each mode described by a cosine function.  This provides a
convenient general purpose approach, but alternative basis functions may allow more accurate representation
of beam images and phase space distributions with fewer modes.  It may be possible, for example, to take
advantage of properties generally expected of the beam (such as approximate symmetries) to construct a
more appropriate basis.  The scope for further development is rather wide, and while the results shown here
are encouraging and demonstrate the value of machine learning for tomographic reconstruction in principle,
more extensive studies would be required to understand the full potential of the technique.

\begin{acknowledgements}

We would like to thank our colleagues in STFC/ASTeC at Daresbury Laboratory
for help and support with various aspects of the simulation and experimental
studies of CLARA.  In particular, we would like to thank Amy Pollard for useful discussions
and advice on machine learning.

This work was supported by the Science and Technology Facilities Council, UK,
through a grant to the Cockcroft Institute.

\end{acknowledgements}


\end{document}